\PassOptionsToPackage{table}{xcolor}
\documentclass[acmsmall]{acmart}
\AtBeginDocument{%
  }

\setcopyright{none}
\copyrightyear{2026}
\acmYear{2026}
\acmDOI{}

\usepackage[english]{babel}
\usepackage{multirow} %
\usepackage{booktabs}
\usepackage{graphicx}
\usepackage{graphicx}
\usepackage{epstopdf}
\usepackage{makecell}

\usepackage{tcolorbox}
\usepackage{amsmath}
\usepackage{graphicx}
\usepackage{subcaption} %
\usepackage{algorithm}
\usepackage{algorithmic}
\usepackage{amsmath}
\usepackage{appendix} %
\AtBeginDocument{%
  }
    
\newtcolorbox{boxK}{
    top=2pt,
    bottom=2pt,
    left=2pt,
    right=2pt,
    boxrule = 0pt,
    toprule = 0pt, %
}
\usepackage{fontawesome5}
\usepackage{adjustbox}
\usepackage{listings}
\usepackage{enumitem}

\definecolor{darkgreen}{rgb}{0.0,0.5,0.0}
\hypersetup{
    colorlinks=true,
    linkcolor=ACMBlue,
    citecolor=ACMBlue,
    urlcolor=ACMBlue,
    pdfborder={0 0 0},
}

\definecolor{code-keyword}{RGB}{139, 0, 0}      %
\definecolor{code-variable}{RGB}{0, 0, 255}      %
\definecolor{code-number}{RGB}{0, 100, 0}       %
\definecolor{code-string}{RGB}{0, 128, 0}       %
\definecolor{code-comment}{RGB}{0, 128, 0}      %
\definecolor{code-operator}{RGB}{100, 100, 100} %
\definecolor{code-bg}{RGB}{255, 255, 255}       %

\newcommand{\code}[1]{\lstinline|#1|}  %

\newcommand{\findingbox}[1]{
    \begin{boxK} 
        \small \faIcon{lightbulb} \textbf{Finding \arabic{findingcounter}:} #1
    \end{boxK}
    \stepcounter{findingcounter}
}
\newcounter{findingcounter}
\begin{document}

\title{Hallucinations of Large Language Models in Cross-Language Code Translation: Phenomena and Mitigation}

\author{Yali Du}
\affiliation{%
  \institution{National Key Laboratory for Novel Software Technology, and the School of Artificial Intelligence, Nanjing University}
  \country{China}}
\email{duyl@lamda.nju.edu.cn}
\orcid{0000-0001-7759-3906}

\author{Yunji Zhang}
\affiliation{%
 \institution{National Key Laboratory for Novel Software Technology, and the School of Artificial Intelligence, Nanjing University}
  \country{China}}
\email{zhangyj@lamda.nju.edu.cn}

\author{Ming Li}
\authornote{Ming Li is the corresponding author.}
\affiliation{%
 \institution{National Key Laboratory for Novel Software Technology, and the School of Artificial Intelligence, Nanjing University}
  \country{China}}
\email{lim@lamda.nju.edu.cn}

\renewcommand{\shortauthors}{Du et al.}

\begin{abstract}
  Code translation converts source code from one programming language to another while preserving its functionality. Although Large Language Models (LLMs) have improved code translation, their outputs still contain hallucinations: generated code that conflicts with the source program, violates target-language facts, or introduces contextually inconsistent dependencies. These failures are especially harmful in translation because small semantic deviations can silently change program behavior. Existing hallucination studies mainly focus on natural-language generation or code generation, leaving the forms, distribution, and mitigation of hallucinations in cross-language code translation insufficiently characterized. This paper presents a systematic empirical study of LLM hallucinations in code translation. Across three benchmark datasets, six language pairs, and seven LLMs, we construct a taxonomy with three major categories: Semantic Consistency Conflicts, Factual Knowledge Conflicts, and Project Context Deviations. We manually annotate 13,919 hallucination cases and analyze how hallucination types vary across models, datasets, and language directions. Guided by the taxonomy, we further propose HaluCodeTran, a mitigation framework that combines BM25-based retrieval of relevant translation examples with taxonomy-aware reflective refinement. 
  The study provides an execution-tested characterization of code-translation hallucinations and a practical baseline for building more reliable LLM-based code migration tools.
\end{abstract}

\ccsdesc[300]{Software and its engineering~Software maintenance tools}

\keywords{Hallucination, Code Translation, Large Language Models}

\maketitle

\section{Introduction}
Code translation, also known as source-to-source compilation or transpilation, involves converting code from one programming language to another while preserving its original functionality. This technique plays a critical role in modern software development. 
For example, many industries spend hundreds of millions of dollars to convert code written in older programming languages (e.g., FORTRAN and COBOL) to newer ones (e.g., Java, C++)~\cite{roziere2020unsupervised}. 
By translating code into modern programming languages, organizations can reduce maintenance costs, improve reliability, and enhance non-functional attributes like security and performance~\cite{pan2024lost}.
Automatic code translation can alleviate the burden of program migration and facilitate the development of software systems, which has drawn sustained attention from neural, LLM-based, and project-scale migration research~\cite{Transcoder,codetran4,du2024joint,pan2024lost,yang2024exploring,macedo2025intertrans}.
Recent advancements in Large Language Models (LLMs) have improved code translation and code-generation capabilities, while also raising new concerns about correctness, calibration, and domain-specific reliability~\cite{yang2024exploring,gu2025domainspecific,spiess2025calibration,shi2025greenllmse}.

However, the translation quality of current LLM-driven approaches is still unsatisfactory, partly because generated programs can contain hallucinations: code fragments that conflict with source-program requirements, violate target-language facts, or introduce contextually inconsistent dependencies. These hallucinations can lead to functional errors, incomprehensible code, deprecated or misused APIs, and security vulnerabilities, undermining the reliability of translated codebases~\cite{pan2024lost,tian2025codehalu,wang2025codeerrors,wang2025libraryevolution,zhuo2026apimisuse}.
Most existing hallucination research focuses on Natural Language Generation (NLG) tasks, such as question answering~\cite{sadat2023delucionqa} and summarization~\cite{rehman2024hallucination,ji2023survey}. Recent software-engineering studies have begun to characterize hallucinations, code-generation errors, and API misuse in LLM-based programming tasks~\cite{liu2024exploring,tian2025codehalu,wang2025codeerrors,lee2025hallucination,zhuo2026apimisuse}, but code translation introduces a different failure surface. Unlike intent-to-code generation, code translation must preserve a structured source program while mapping syntax, types, libraries, APIs, and idioms into a different target language. A generated program may compile and still violate the source semantics, while a superficially small API or data-flow mismatch can completely change behavior. This makes hallucination in code translation a distinct empirical object rather than a direct instance of natural-language or code-generation hallucination.

To address this gap, we conduct an empirical study focused on hallucinations in cross-language code translation. Through open coding and empirical validation, we develop a taxonomy of code translation hallucinations, categorizing errors into three main types:
(1) Semantic Consistency Conflicts (e.g., functional requirement mismatches, data inconsistencies),
(2) Factual Knowledge Conflicts (e.g., syntax errors, incorrect library/API usage),
(3) Project Context Deviations (e.g., dead code, dependency conflicts). We instantiate this taxonomy on outputs from seven LLMs, including GPT-4o~\cite{hurst2024gpto} and Qwen 2.5-Coder~\cite{yang2024qwen2.5}, over three widely used code translation datasets: CodeNet~\cite{codenet}, Avatar~\cite{codetran4}, and Transcoder~\cite{Transcoder}. The resulting annotations provide a basis for analyzing which hallucination categories dominate code translation failures and which model/language-pair conditions make them more likely.

Guided by the taxonomy, we propose HaluCodeTran, a mitigation framework that combines retrieval-based contextual grounding with reflective iterative refinement. The retrieval component provides relevant source-target translation examples to reduce factual and idiomatic uncertainty, while the reflection component asks the model to inspect the intermediate translation against the source program and taxonomy categories before producing a revision. On CodeNet, HaluCodeTran improves computational accuracy by up to 23 percentage points and reduces hallucination cases by up to 32\% compared with vanilla translation, outperforming Chain-of-Thought, Chain-of-Verification, and Retrieval-Augmentation in average accuracy for the evaluated models.

We summarize our contributions as follows:
\begin{itemize}
    \item We define hallucination in code translation and establish a taxonomy that distinguishes semantic consistency conflicts, factual knowledge conflicts, and project context deviations.
    \item We construct a manually labeled hallucination dataset covering 13,919 cases from seven LLMs, three datasets, and six translation directions, enabling fine-grained analysis across models and language pairs.
    \item We propose HaluCodeTran, a taxonomy-aware mitigation framework that combines retrieval-augmented examples with reflective refinement, and evaluate it against representative prompting and retrieval baselines.
\end{itemize}

\section{Related Works}
\subsection{LLM-Powered Code Translation}
Automated code translation facilitates the migration of codebases across programming languages and can reduce manual migration effort. Early work focused on rule-based approaches~\cite{rule1,rule2,rule3,rule4} and phrase-based statistical machine translation techniques~\cite{KaraivanovRV14,NguyenNN15}, which laid the groundwork for subsequent neural and LLM-based approaches~\cite{pscpt,codetran4,must,du2024joint,macedo2025intertrans}.
Recently, large language models (LLMs) with hundreds of billions of parameters have demonstrated impressive capabilities across natural language and code tasks~\cite{hurst2024gpto,hui2024qwen2}. 
General-purpose LLMs~\cite{yang2024qwen2.5,grattafiori2024llama,mistral7b,abdin2024phi} and code-specialized LLMs~\cite{codellama,team2024codegemma,li2023starcoder,lozhkov2024starcoder2} have shown strong performance when trained on large code corpora, such as GPT~\cite{tang2023chatgpt}, LLaMA~\cite{grattafiori2024llama}, Qwen~\cite{hui2024qwen2, yang2024qwen2.5}, Gemma~\cite{team2024gemma}, DeepSeek-Coder~\cite{guo2024deepseek,zhu2024deepseek}, etc.

The LLM-powered code translation has achieved significant progress in recent years~\cite{saha2024specification,macedo2024exploring,yang2024exploring,shiraishi2024context,tao2024unraveling,dhruv2024leveraging,zhang2024scalable,yuan2024transagent,jana2024cotran,luo2024bridging,macedo2025intertrans}.
They enhance the code translation performance of large language models through technical approaches such as prompt engineering optimization and domain-specific fine-tuning. 
For example, by augmenting generation with in-context learning, the code structure (e.g., Abstract Syntax Tree) knowledge is leveraged into large language models to improve understanding of cross-language semantic alignment~\cite{du2025post}. Meanwhile, domain-specific fine-tuning on specialized codebases enables models to better capture niche terminology and idiomatic expressions in specific fields~\cite{dhruv2024leveraging}. Additionally, techniques like multi-model ensemble and interactive feedback mechanisms are employed to enhance the accuracy and practicality of code translation~\cite{yang2024exploring}. 

\subsection{Hallucination of Large Language Models}

In the field of natural language processing (NLP), hallucination refers to the phenomenon where a language model generates content that is inconsistent with the input, semantically meaningless, or factually incorrect. With the rapid development of LLMs, research on hallucination has attracted significant attention across various domains~\cite{ji2023survey,huang2025survey,lavrinovics2025knowledge,lauscher2025much}.
In the community of AI for Software Engineering, existing investigations primarily focus on hallucinations, calibration, preference alignment, and error characteristics within code generation and completion~\cite{liu2024exploring,tian2025codehalu,spiess2025calibration,wang2025codeerrors,weyssow2026codeultrafeedback,zhuo2026apimisuse}. 
However, a fundamental distinction exists between code generation and code translation. While code generation requires the model to understand ambiguous natural language intents and write programs, code translation demands a rigorous understanding of structured source code and precise cross-lingual semantic mapping. This shift from ``intent-to-code'' to ``code-to-code'' alignment introduces distinct hallucination patterns, such as syntax-level interference, cross-language API mismatches, and subtle source-target semantic divergence. Existing code-hallucination studies provide useful definitions and evaluation ideas, but they do not directly answer which hallucination types dominate execution-tested code translation failures or how these types vary across source-target language pairs. Our study therefore scopes hallucination to cross-language translation and evaluates it through compiler/runtime behavior, source-target comparison, and manual taxonomy annotation.

\subsection{Hallucination Mitigation via Prompt Engineering}
Recent research has extensively explored prompt engineering as a cost-effective paradigm to suppress hallucinations and improve code-generation reliability by optimizing the model’s reasoning paths and grounding its outputs in verifiable contexts~\cite{wei2022chain,khojah2025promptprogramming}.
First, Reasoning-guided Mitigation, exemplified by Chain-of-Thought (CoT) prompting~\cite{wei2022chain}, encourages LLMs to generate intermediate logical steps, which facilitates the decomposition of complex source code structures. However, CoT is often susceptible to ``logical drifting'', where a minor hallucination in the early reasoning stage cascades into a significant functional requirement mismatch in the final translated code.
Second, Self-Verification Mechanisms, such as the Chain-of-Verification (CoVe) framework~\cite{chainofverification} and recent program-analysis/backtracking strategies~\cite{jiang2025rocode}, introduce a ``generate-then-verify'' pipeline to rectify factual errors through independent query-answering or code-level analysis. While effective for general knowledge and some code-generation settings, these mechanisms can still fail to recognize subtle API knowledge conflicts across heterogeneous programming languages~\cite{zhuo2026apimisuse}.
Third, Knowledge-Augmented Approaches, particularly Retrieval-Augmented Generation (RAG)~\cite{lewis2020retrieval,zhao2023felm,du2025post}, mitigate hallucinations by providing external ``ground-truth'' snippets from documentation or project-specific repositories. Despite its success in reducing project context deviations, RAG remains inherently limited by the quality of the retrieval corpus, and recent TSE evidence shows that code datasets can contain cross-dataset duplication and leakage that complicate LLM evaluation~\cite{lopez2025leakage}.

These lines of work leave two gaps for code translation. First, code translation studies primarily report end-to-end correctness metrics, but they rarely decompose failed translations into hallucination categories that explain what kind of unsupported content the model introduced. Second, hallucination mitigation methods are usually designed as generic prompting or retrieval strategies, while code translation requires feedback that is aligned with source-target semantic preservation, target-language factual knowledge, and project-context consistency. To address these gaps, our work studies hallucination phenomena in code translation and proposes a mitigation approach grounded in the observed taxonomy.

\begin{figure*}[t]
\centering
\includegraphics[width=\linewidth]{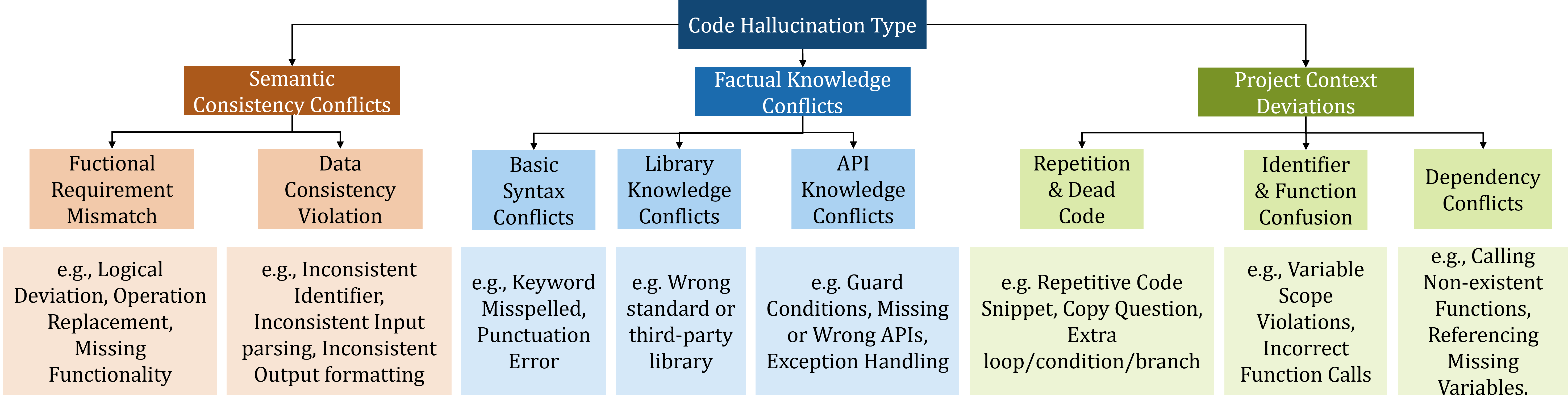}
\caption{Taxonomy of Hallucinations in LLM-based Code Translation}
\label{fig:taxonomy}
\end{figure*}

\begin{figure*}[!ht]
    \centering  
    
    \begin{subfigure}[b]{0.31\linewidth}
        \includegraphics[width=\textwidth]{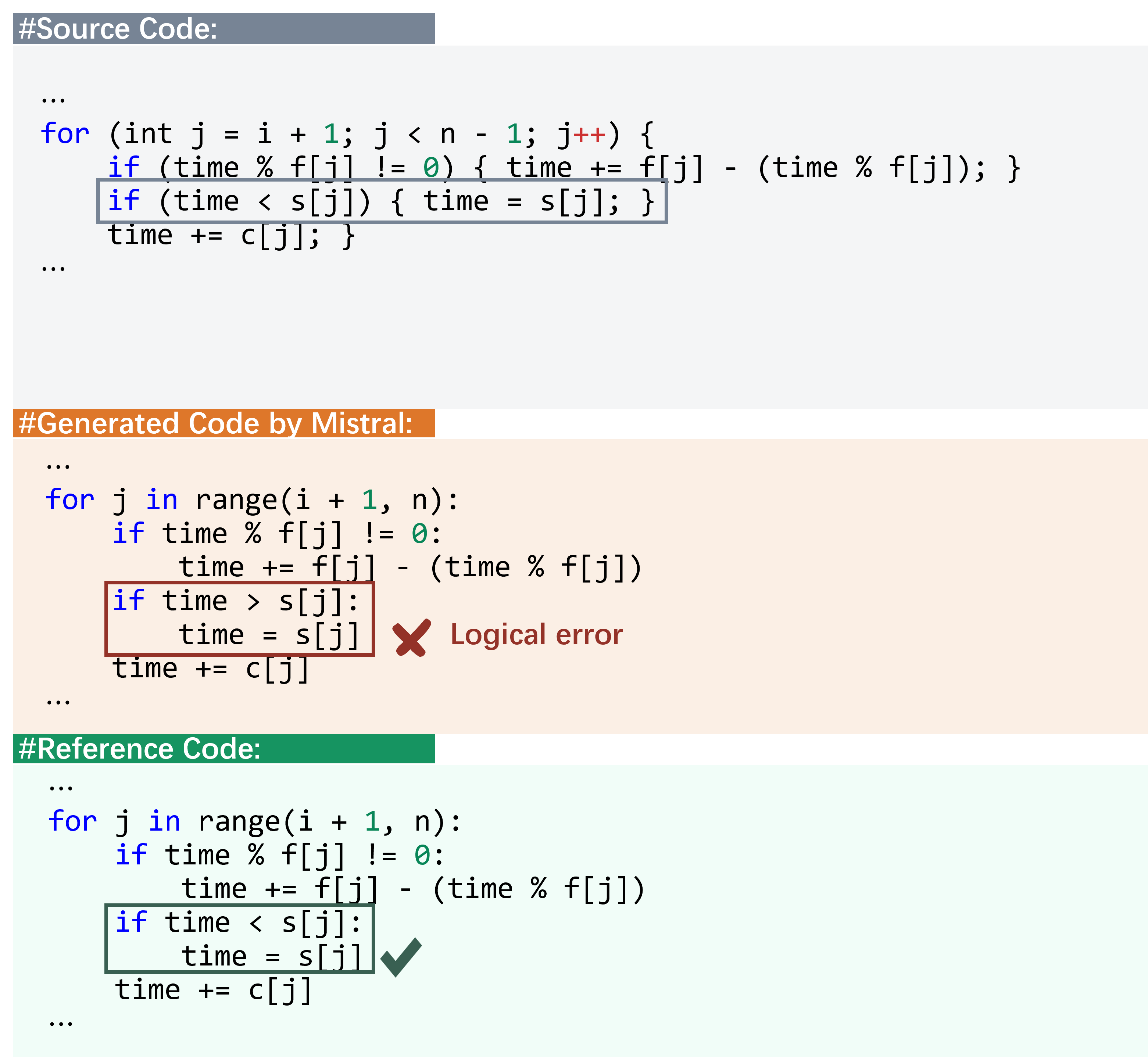}
        \caption{Functional Requirement Mismatch}
        \label{fig:frm}  %
    \end{subfigure}\hfill
    \begin{subfigure}[b]{0.31\linewidth}
        \includegraphics[width=\textwidth]{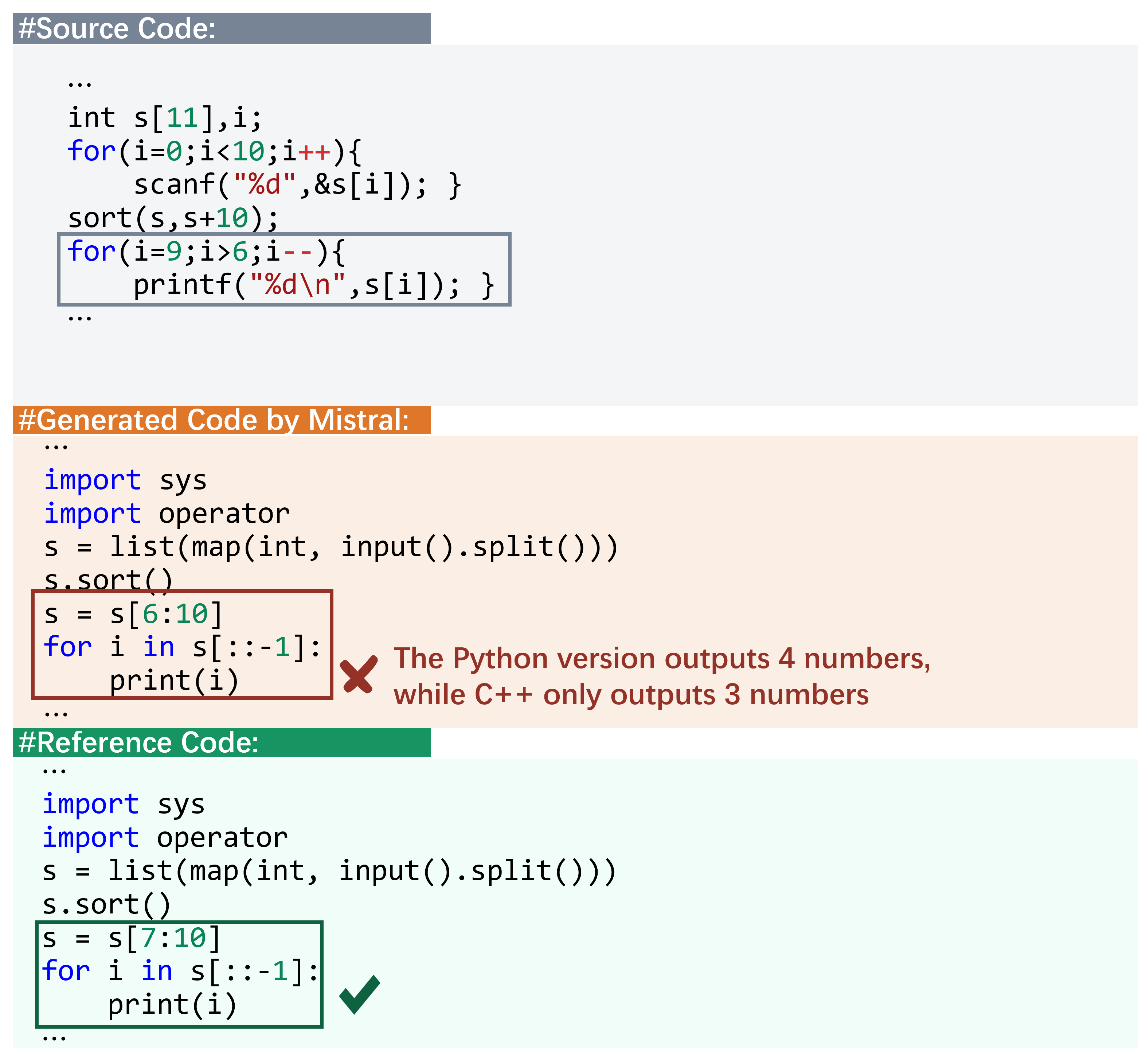}
        \caption{Data Consistency Violation}
        \label{fig:dcv}
    \end{subfigure}\hfill
    \begin{subfigure}[b]{0.31\linewidth}
        \includegraphics[width=\textwidth]{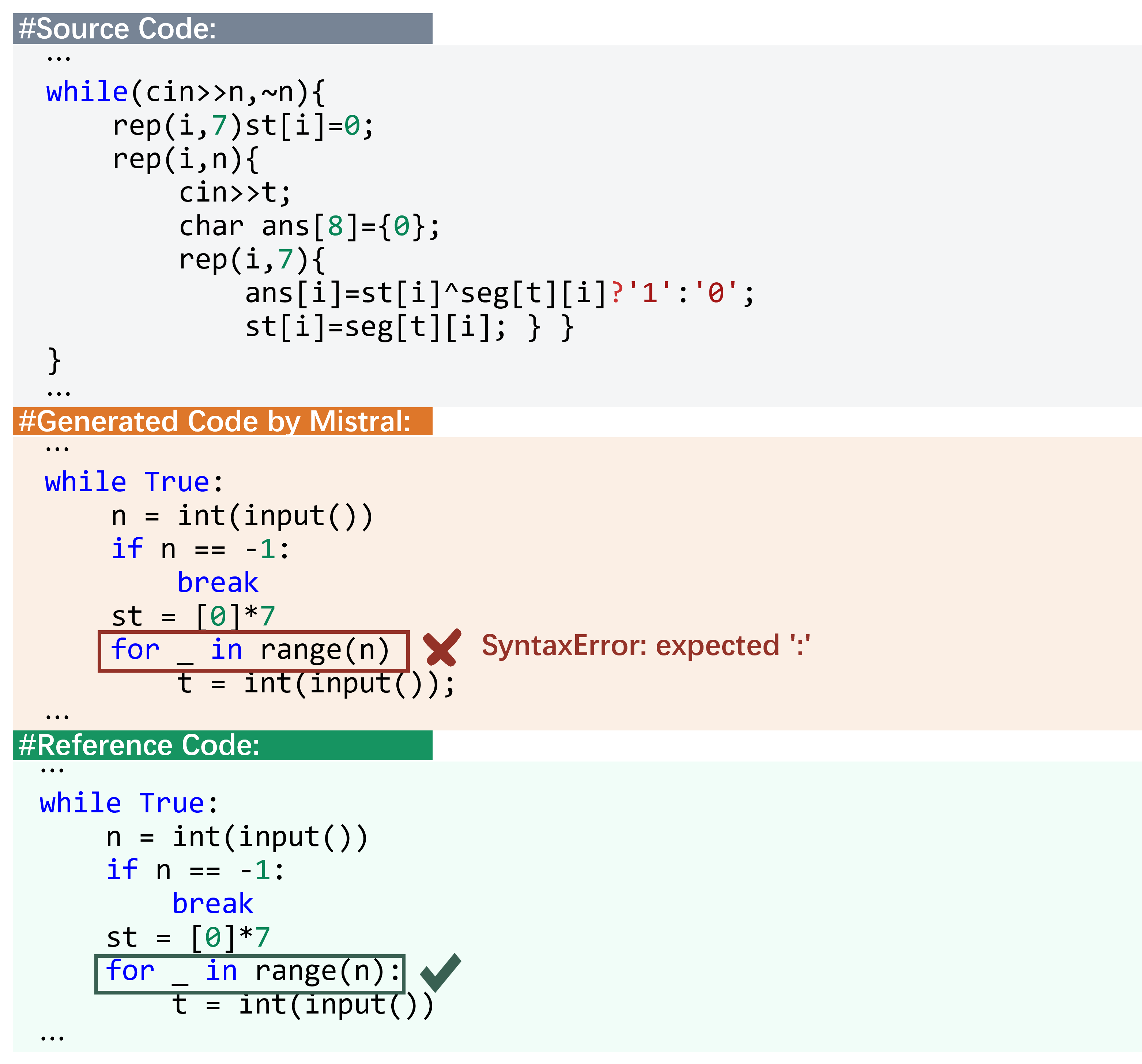}
        \caption{Basic Syntax Error}
        \label{fig:bsc}
    \end{subfigure}
    
    \begin{subfigure}[b]{0.31\linewidth}
        \includegraphics[width=\textwidth]{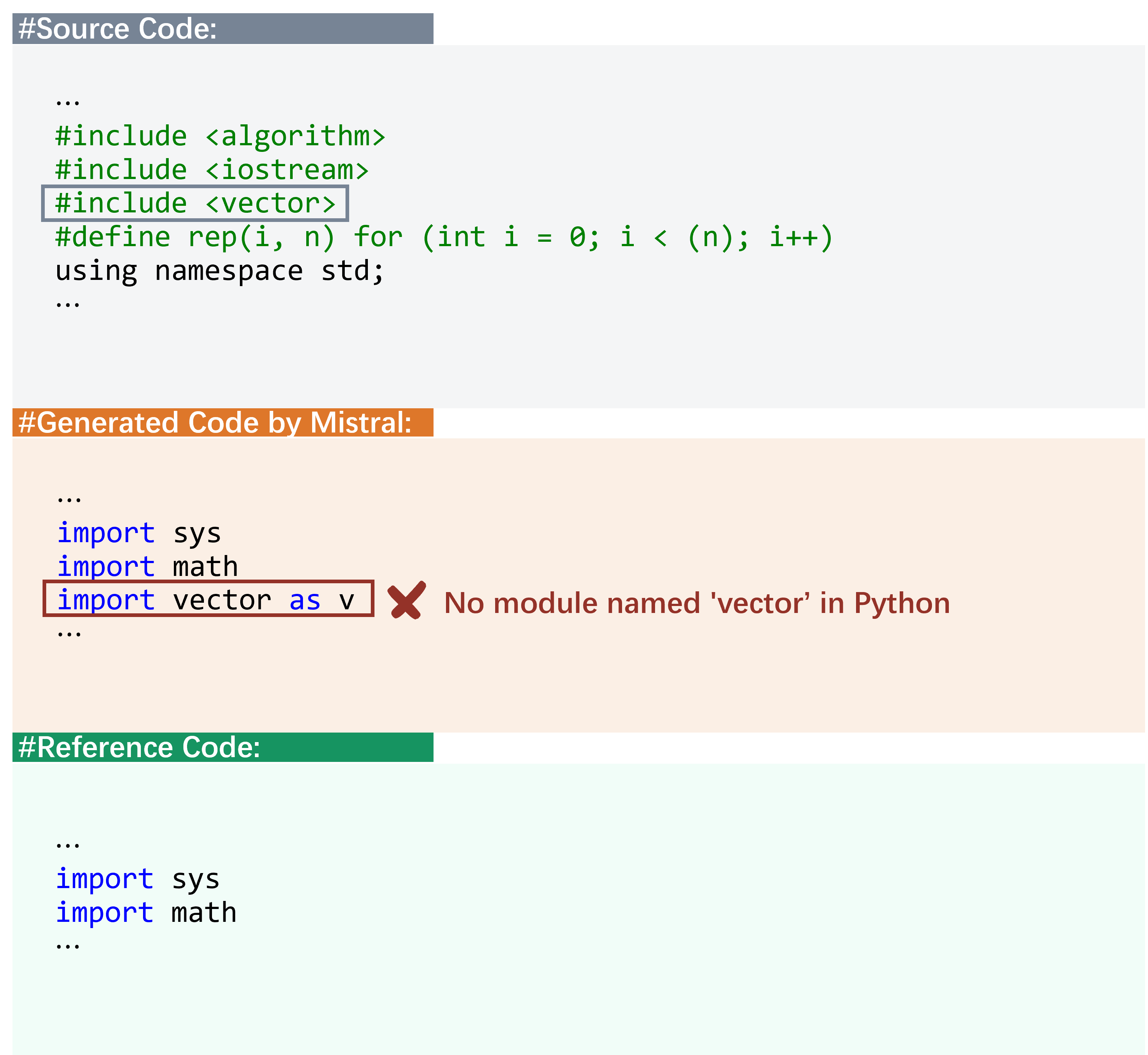}
        \caption{Library Knowledge Conflicts}
        \label{fig:lkc}
    \end{subfigure}\hfill
    \begin{subfigure}[b]{0.31\linewidth}
        \includegraphics[width=\textwidth]{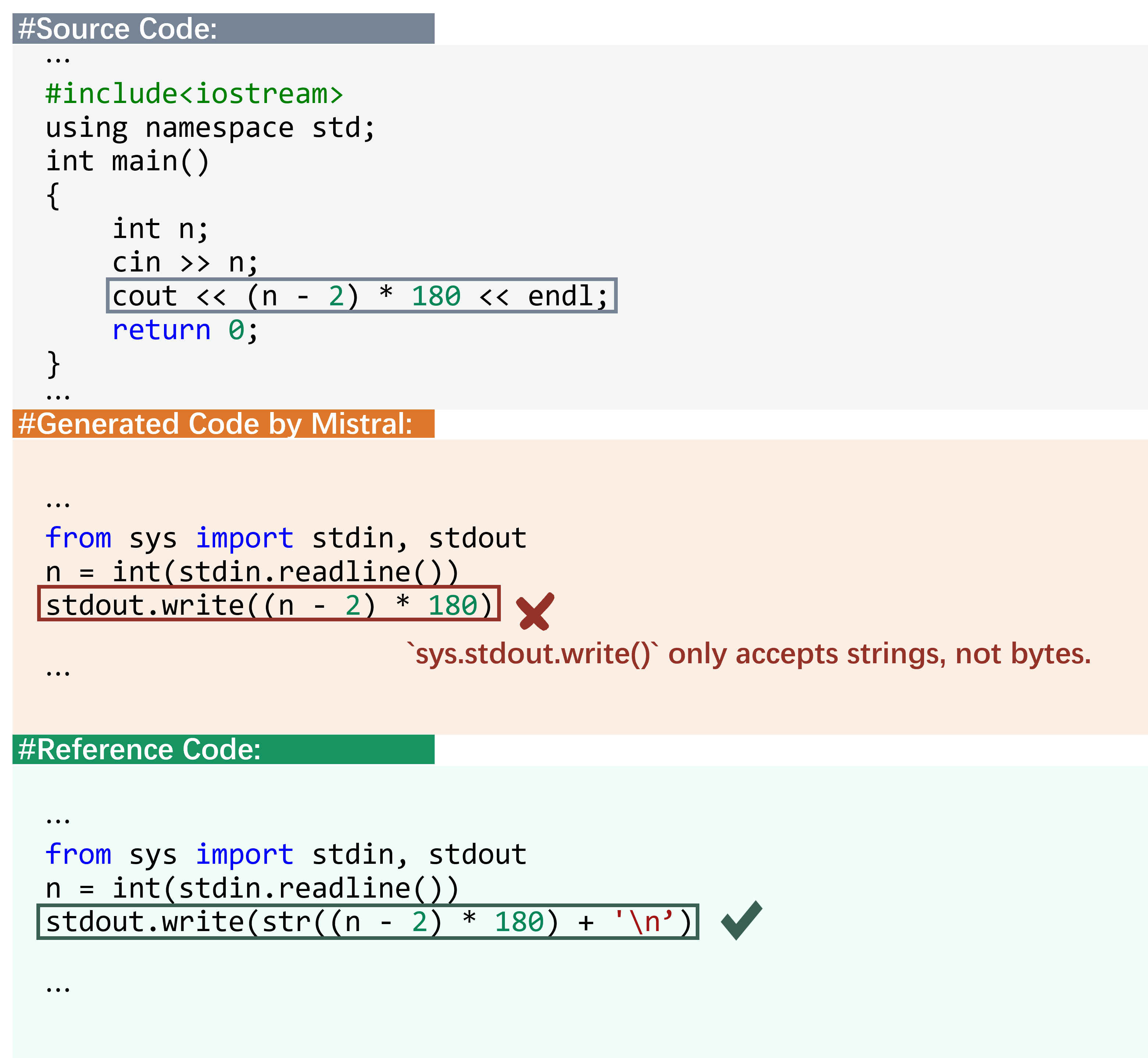}
        \caption{API Knowledge Conflicts}
        \label{fig:akc}
    \end{subfigure}\hfill
    \begin{subfigure}[b]{0.31\linewidth}
        \includegraphics[width=\textwidth]{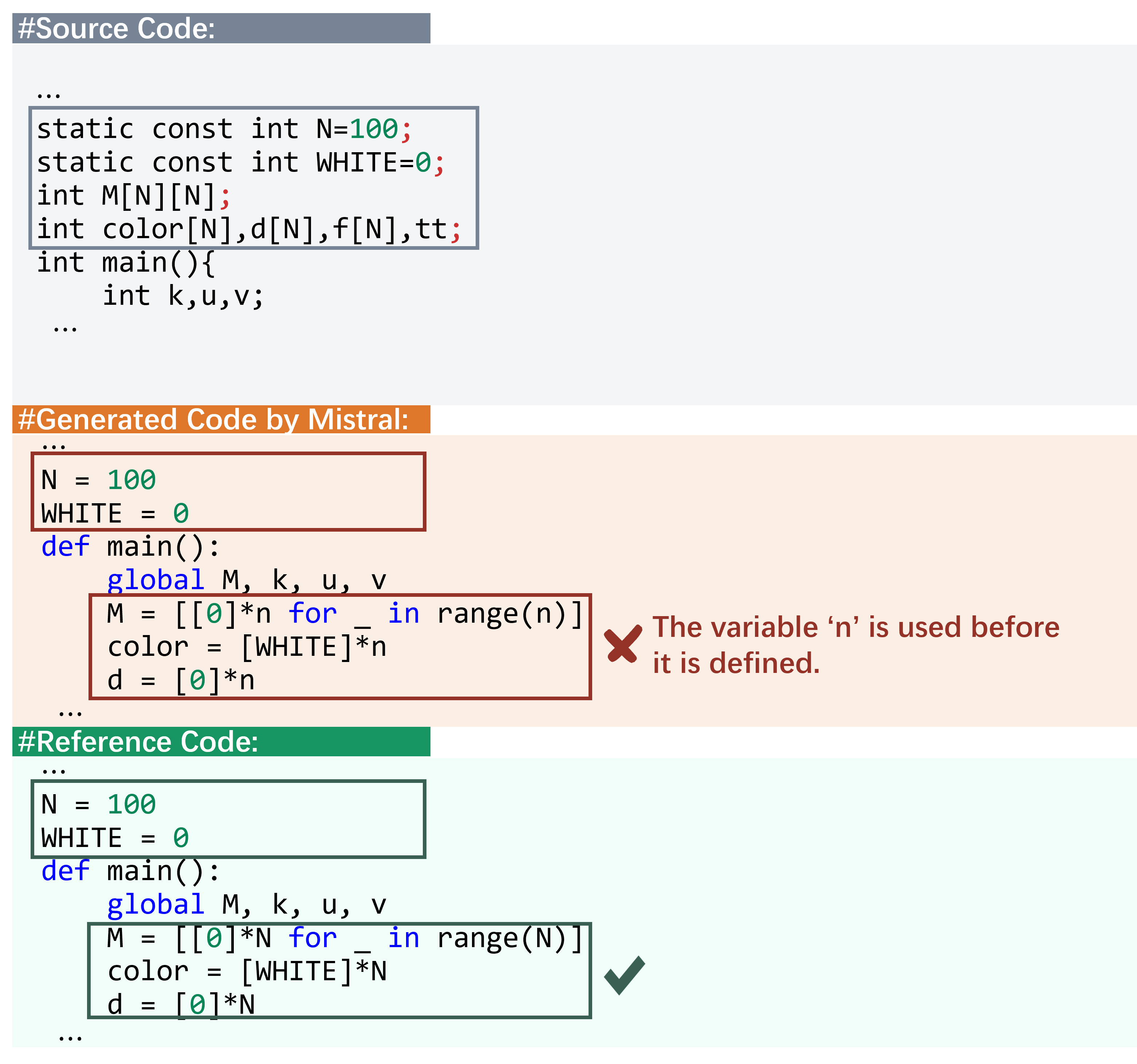}
        \caption{Dependency Conflicts}
        \label{fig:dc}
    \end{subfigure}
    
    \begin{subfigure}[b]{0.45\linewidth}
        \includegraphics[width=\textwidth]{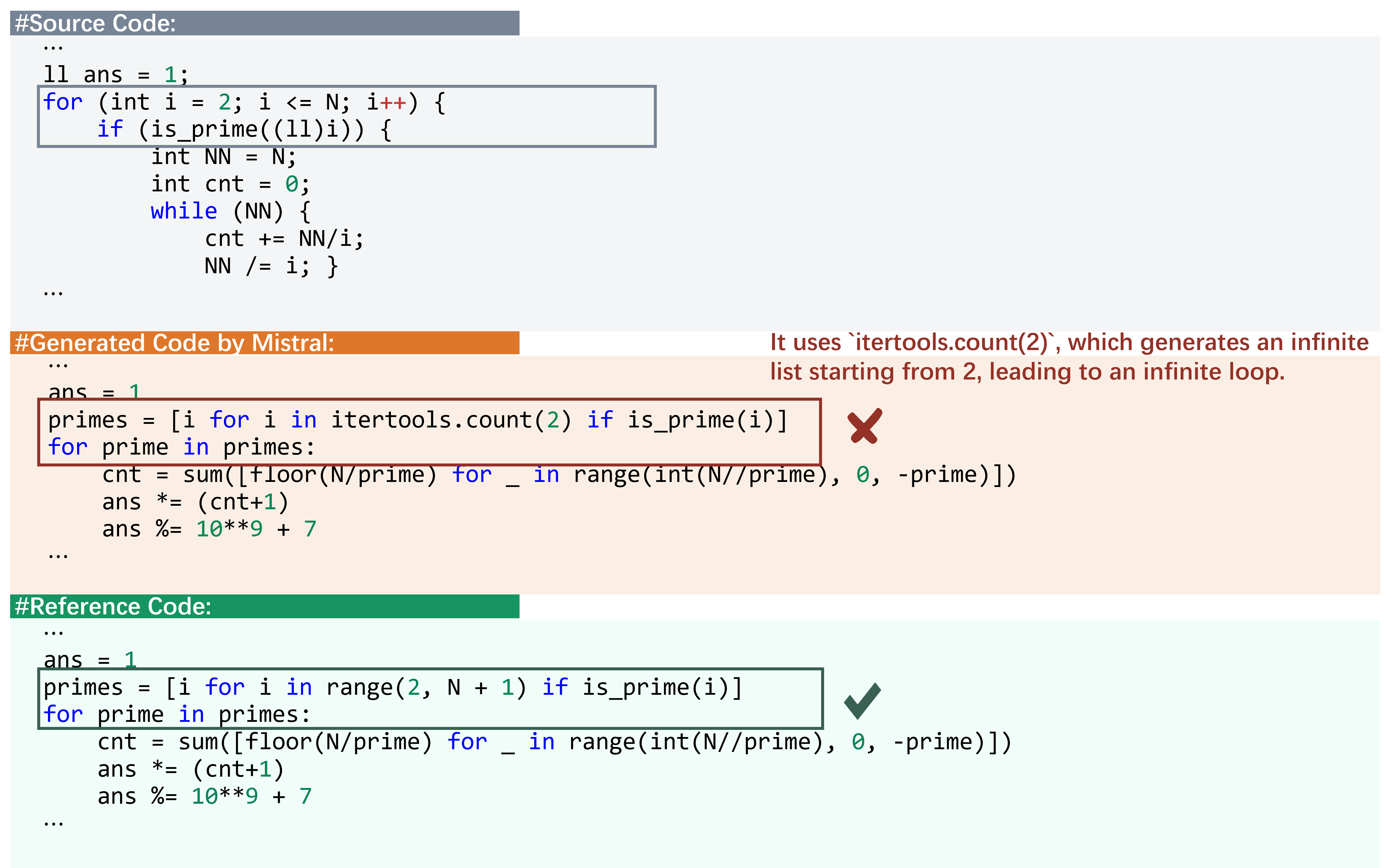}
        \caption{Repetition \& Dead Code}
        \label{fig:rdc}  %
    \end{subfigure}\hfill
    \begin{subfigure}[b]{0.45\linewidth}
        \includegraphics[width=\textwidth]{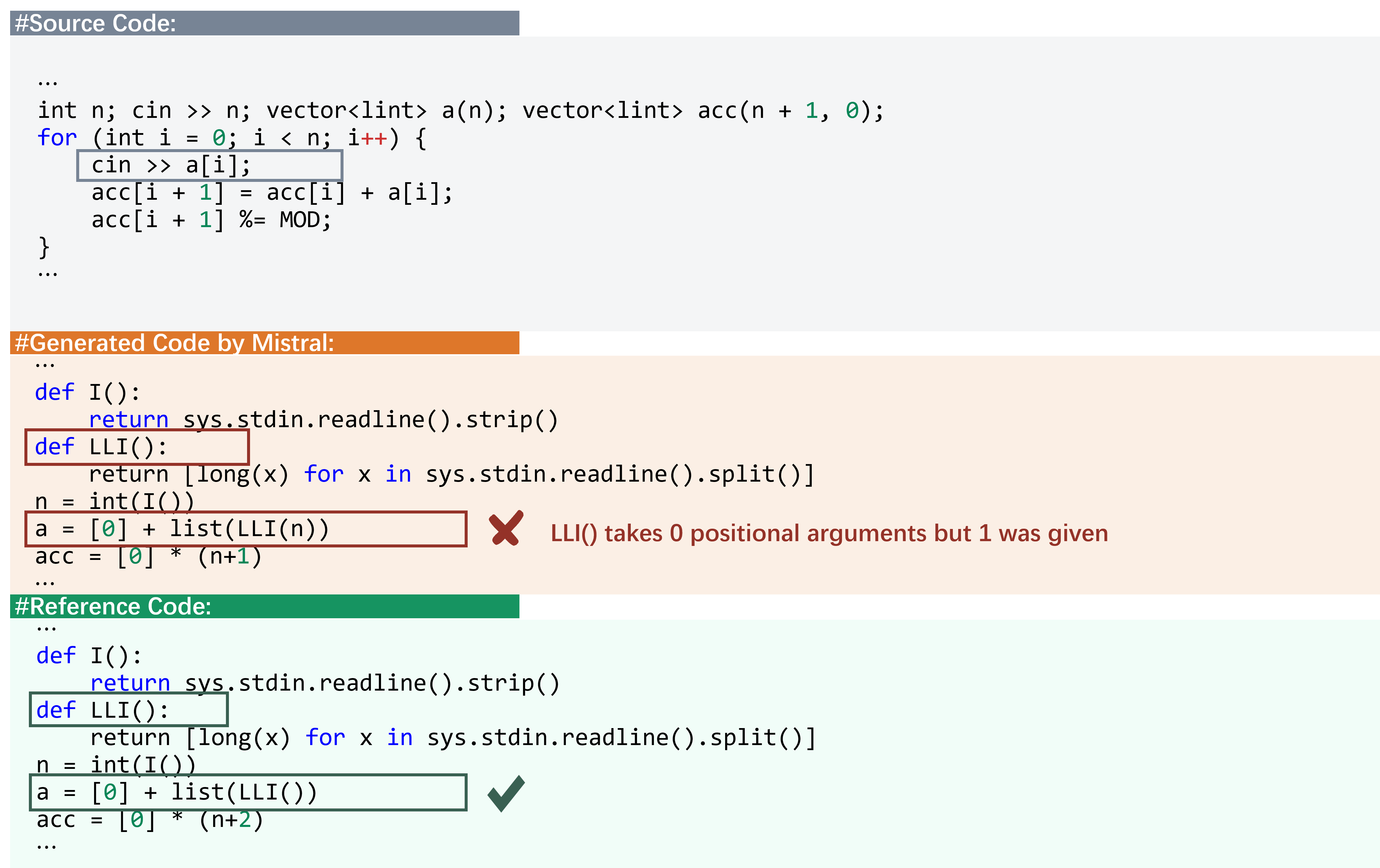}
        \caption{Identifier \& Function Confusion}
        \label{fig:ifc}
    \end{subfigure}
    
    \caption{Examples of Different Hallucination Types in LLM-Powered Code Translation}
    \label{fig:hallucination_examples}
\end{figure*}

\section{Taxonomy of Hallucination}
We define a code-translation hallucination as a generated target-language program element that is not justified by the source program or target-language semantics and that can affect compilation, execution, maintainability, or dependency consistency. This definition separates hallucination from ordinary translation diversity. A target program is not labeled as hallucinated merely because it differs textually from the reference implementation: semantically equivalent rewrites, idiomatic target-language substitutions, and harmless formatting differences are treated as valid variants. In contrast, we label a generated fragment as hallucinated when test execution, compiler/runtime diagnostics, or manual source-target comparison shows that the fragment introduces unsupported behavior, invalid language/library knowledge, or inconsistent project context.

To construct the taxonomy, two authors first performed open coding on LLM-generated translations from the benchmark setting described in Section~\ref{subsectioncomparision}. We employ Qwen 2.5-Coder~\cite{yang2024qwen2.5}, DeepSeek-Coder-V2~\cite{zhu2024deepseek}, Llama 3.1~\cite{grattafiori2024llama}, Gemma 2~\cite{team2024gemma}, Mistral~\cite{mistral7b}, Phi 4~\cite{abdin2024phi}, and GPT-4o~\cite{hurst2024gpto} on CodeNet~\cite{codenet}, Avatar~\cite{codetran4}, and Transcoder~\cite{Transcoder}, covering C++$\rightarrow$Java, C++$\rightarrow$Python, Java$\rightarrow$C++, Java$\rightarrow$Python, Python$\rightarrow$C++, and Python$\rightarrow$Java. Each generated program is compiled or executed in the corresponding target-language environment. The annotators then inspect the generated code, compiler/runtime messages, source program, reference solution, and available test results to identify the concrete erroneous fragment and assign one or more hallucination labels. This process follows open-coding practice and is informed by hallucination taxonomies from natural-language and code-generation studies~\cite{zhang2024llm,tian2025codehalu}.

The annotation protocol is multi-label because a single failed translation can contain more than one hallucinated fragment or one fragment can exhibit multiple failure modes. For example, an undefined function call can simultaneously reflect identifier confusion and dependency conflict if the call is inconsistent with both local scope and required imports. When annotators encounter a case not covered by the current codebook, they record the evidence, propose a provisional label, and discuss whether the case should be merged into an existing category or used to refine the taxonomy. Disagreements are resolved through joint inspection of the source program, generated program, execution result, and category definition. 

Figure~\ref{fig:taxonomy} summarizes the resulting taxonomy. It contains three major categories and eight subcategories. In total, we label 13,919 hallucination cases generated by seven LLMs. The following subsections define each category and provide representative examples from the hallucination dataset.
\subsection{Semantic Consistency Conflicts}\label{scc}
Semantic consistency conflicts occur when the translated code fails to preserve the intended behavior or semantics of the source code.

\subsubsection{Functional Requirement Mismatch (FRM)} \label{frm}
Functional requirement mismatch refers to the generated code diverging from the intended logic or functional requirements.
As shown in Figure~\ref{fig:hallucination_examples}(\subref{fig:frm}), the source C++ code has a conditional judgment \lstinline|if (time < s[j])|, which means that when time is less than \lstinline|s[j]|, the value of time should be updated to \lstinline|s[j]|. However, the generated Python code wrongly writes the condition as \lstinline|if time > s[j]|. This incorrect logical condition is a misinterpretation of the original code's intention, deviating from the correct logic that the source code wants to implement, thus resulting in a logical error.

\subsubsection{Data Consistency Violation (DCV)} \label{dcv}
Data consistency violation occurs when variables, inputs, or outputs become inconsistent during translation.
As shown in Figure~\ref{fig:hallucination_examples}(\subref{fig:dcv}), the LLM misuses multiple assignments by incorrectly incorporating the augmented assignment \lstinline|sum += num[i] * m[i]| into a multiple-assignment statement. The augmented assignment is not valid as the left-hand side of the multiple assignment must be a legitimate assignable target, and \lstinline|sum += num[i] * m[i]| does not meet this requirement. 

\subsection{Factual Knowledge Conflicts}\label{fkc}
Factual knowledge conflicts arise when the translated code violates the basic syntax or demonstrates misunderstandings of libraries or APIs because of the limitations of the knowledge of the corresponding programming languages.
\subsubsection{Basic Syntax Conflicts (BSC)}\label{bsc}
Basic Syntax Conflicts involve syntax-related errors that render the translated code incorrect or uncompileable.
In the Figure~\ref{fig:hallucination_examples}(\subref{fig:bsc}), in Python code, a colon is required at the end of the for loop header to denote the start of the loop body. However, the generated code has \lstinline|for _ in range(n)| without the necessary colon, leading to a SyntaxError as the interpreter expects a colon to properly structure the statement.

\subsubsection{Library Knowledge Conflicts (LKC)}\label{lkc}
Library knowledge conflicts occur when the translation uses incorrect or non-existent standard or third-party libraries. As shown in Figure~\ref{fig:hallucination_examples}(\subref{fig:lkc}), there is no standard module called vector in Python, yet the code generation wrongly assumes its existence and tries to import it, mimicking the \lstinline|#include <vector>| statement from the C++ source code. This is a clear case of the code generation tool generating incorrect code based on a misinterpretation or over-generalization of the source code's library inclusion mechanism, failing to account for the differences in module availability between C++ and Python.

\subsubsection{API Knowledge Conflicts (AKC)}\label{akc}
API knowledge conflicts involve incorrect usage of APIs, including function names, parameters, or control logic. As shown in Figure~\ref{fig:hallucination_examples}(\subref{fig:akc}), the generated code attempts to directly pass a numerical result \lstinline|(n - 2) * 180| to \lstinline|sys.stdout.write()|, ignoring the fact that this function only accepts strings. This is a misinterpretation of how output operations work in the API of Python.

\subsection{Project Context Deviations}\label{pcd}
Project context deviations happen when the generated code fails to align with the broader context of the project, such as naming conventions, dependencies, or logic flow.

\subsubsection{Repetition \& Dead Code (RDC)}\label{rdc}
Repetition or dead code issues include unnecessary or unreachable code introduced in translation.
In Figure~\ref{fig:hallucination_examples}(\subref{fig:rdc}), since \lstinline|itertools.count(2)| creates an infinite list starting from 2, it leads to an infinite loop when iterating over the resulting primes list. The generated code wrongly assumes an infinite sequence is appropriate, resulting in incorrect and non-terminating behavior.

\subsubsection{Identifier \& Function Confusion (IFC)}\label{ifc}
Identifier \& function confusion refers to hallucinations where the generated code misuses variables, functions, or instances that are defined within the local project context (rather than external APIs). These errors often indicate a failure to track or understand in-project definitions, scopes, or naming.
As shown in Figure~\ref{fig:hallucination_examples}(\subref{fig:ifc}), the generated code attempts to call \lstinline|LLI(n)| while the function \lstinline|LLI| is defined to take no positional arguments. This misinterpretation of the function's parameter requirements leads to a runtime error stating that ``\lstinline|LLI()| takes 0 positional arguments but 1 was given''. It shows that the code generation process has an incorrect understanding of how the function should be called, creating an inconsistent and erroneous implementation compared to the intended logic.

\subsubsection{Dependency Conflicts (DC)}\label{dc}
Dependency conflicts occur when the generated code refers to functions, attributes, variables, or APIs that do not exist in the project context. Due to the inability of LLMs to access the entire project context, they often hallucinate dependencies by calling non-existent functions, referencing missing variables or attributes, etc.
As shown in Figure~\ref{fig:hallucination_examples}(\subref{fig:dc}), in the code generated by Mistral, there is an error of using variables before they are defined. In Python, a variable must be defined and assigned a value before it can be used. In the main function of the generated code, in statements like \lstinline|M = [[0]* N, for _ in range(N)]|, \lstinline|color = [WHITE]*N, d = [0]*N, f = [0]*N|, the non-existing variables \lstinline|N| is used within the program.

\section{Evaluation of Hallucination}
\subsection{Dataset and Studied LLMs}
To evaluate the hallucination in code translation, we perform an extensive experiment on three well-known datasets of code translation: CodeNet~\cite{codenet}, Avatar~\cite{codetran4}, and Transcoder~\cite{Transcoder}. The metric to measure the performance of the LLMs in code translation is computational accuracy, proposed by Transcoder~\cite{Transcoder}.

Seven advanced LLMs are evaluated: GPT-4o, four open-source LLMs (Llama 3.1, Gemma 2, Mistral, and Phi 4), and two recent code LLMs (Qwen 2.5-coder, DeepSeek-Coder-V2).

\begin{itemize}[leftmargin=*]
    \item \includegraphics[width=.03\linewidth]{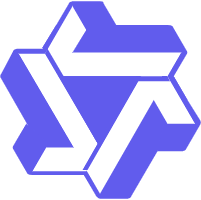} Qwen 2.5-Coder~\cite{hui2024qwen2}: is a code-specific model, which is built upon the Qwen2.5 architecture and continues to be pretrained on a vast corpus of over 5.5 trillion tokens. We use the Qwen 2.5-coder-7B model. 
    \item \includegraphics[width=.03\linewidth]{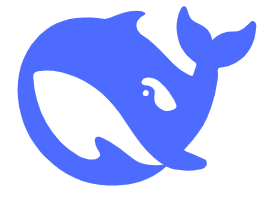} DeepSeek-Coder-V2~\cite{zhu2024deepseek}: is an open-source Mixture-of-Experts (MoE) code language model further pre-trained from an intermediate checkpoint of DeepSeek-V2. We use the DeepSeek-Coder-V2-16B model.
    \item \includegraphics[width=.03\linewidth]{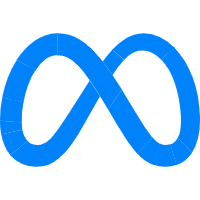} Llama 3.1~\cite{grattafiori2024llama}: is a new state-of-the-art model from Meta available in 8B, 70B, and 405B parameter sizes that natively support multilinguality, coding, reasoning, and tool usage. We use the Llama-3.1-8B model.
    \item \includegraphics[width=.03\linewidth]{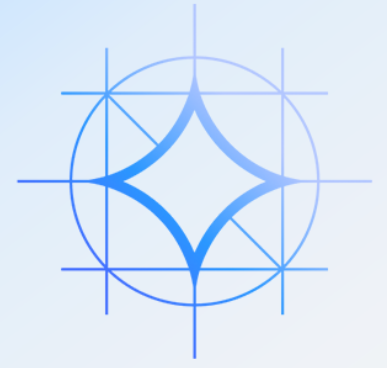} Gemma 2~\cite{team2024gemma}: a new addition to the Gemma family of lightweight models, ranging in scale from 2 billion to 27 billion parameters for text and code. We use the Gemma 2-9B model.
    \item \includegraphics[width=.03\linewidth]{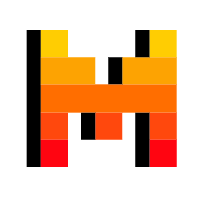} Mistral~\cite{mistral7b}: is a 7B parameter model released by Mistral AI, which is available in both instruction following and text completion.
    \item \includegraphics[width=.03\linewidth]{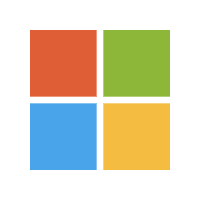} Phi 4~\cite{abdin2024phi}: is a 14-billion parameter language model developed with a training recipe that is centrally focused on data quality.
    \item \includegraphics[width=.025\linewidth]{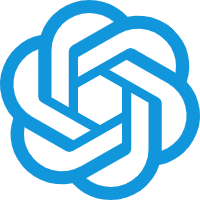} GPT-4o~\cite{hurst2024gpto}: is an advanced generative AI model developed by OpenAI. Although not explicitly trained for code translation or code generation, they demonstrate notable performance in this domain.
\end{itemize}

For each dataset and translation direction, we use the official source programs and available test cases to evaluate generated target programs. A generated program is counted as compilable only when it passes the target-language compilation or syntax-checking stage without manual repair. A generated program is counted as computationally correct only when it produces the same outputs as the reference program on the provided test suite. We do not use test outputs, reference explanations, or ground-truth target code as feedback to the LLM during vanilla translation or reflective refinement; these artifacts are used only for offline evaluation and annotation.

\subsection{Evaluation Metrics}
For translation correctness, we follow most previous studies~\cite{roziere2020unsupervised,Transcoder,yang2024exploring,du2025post} to measure two main aspects, including the Compilation Success Rate (CSR) and Computational Accuracy (CA).

\textbf{Compilation Success Rate (CSR):} It computes the ratio of samples that can be successfully compiled after translation. 
\begin{equation}
    CSR = \frac{\sum^{N_{x}}_{k=1}cs(\hat{y_{k}})}{N_{x}}, \ \textbf{where} \ cs(\hat{y_{k}}) = \left\{\begin{matrix}
1 & Compile(\hat{y_{k}}) \to success\\
0 & Compile(\hat{y_{k}})  \to error\\
\end{matrix}\right. 
\end{equation}
where $N_{x}$ denotes the total number of samples, $\hat{y_{k}}$ denotes the $k$-th translated sample via a certain LLM. \textit{Compile($\cdot$)} denotes compiling samples with their corresponding compilation programs, such as \textit{javac} for Java while \textit{g++} for C++, and its results are either \textit{success} or \textit{error}.

\textbf{Computational Accuracy (CA):} It computes the ratio that the translated programs can produce the same execution result as the ground truths, given the same inputs. 
\begin{equation}
CA = \frac{\sum^{N_{x}}_{k=1}ca(y_{k},\hat{y_{k}})}{N_{x}}, \ \textbf{where} \ ca(y_{k},\hat{y_{k}}) = \left\{\begin{matrix}
1 & Exec_{k}(y_k)=Exec_{k}(\hat{y_{k}})\\
0 & Exec_{k}(y_k)\neq Exec_{k}(\hat{y_{k}}) \\
\end{matrix}\right. \label{eq2}
\end{equation}
where 
$N_{x}$ denotes the number of samples for translation, 
$y_k$ denotes the ground truth of the $k$-th
sample, and $\hat{y_{k}}$ denotes the translated program via a certain LLM for the $k$-th sample. Besides,
$Exec_{k}(\cdot)$ denotes the execution result of a program with the test suite of the $k$-th sample.

When interpreting CA, we distinguish absolute percentage-point changes from relative percentage changes. For example, an improvement from 33.58\% to 50.33\% is reported as a 16.75 percentage-point gain. This convention avoids conflating relative error reduction with absolute translation accuracy improvement.

\section{Evaluation Results}
\subsection{RQ1: What is the performance of the Vanilla LLMs on code translation?}
\label{subsectioncomparision}

This RQ establishes the foundational baseline for the entire empirical study, which is essential for objectively measuring the native code translation capabilities of state-of-the-art LLMs via core metrics like CSR and CA. It quantifies performance disparities across models and language pairs, and provides a measurable reference for attributing unsatisfactory translation results to hallucination issues, laying the empirical groundwork for all subsequent hallucination-related analyses.

Table~\ref{comparision} summarizes the computation accuracy of code translation across three datasets (CodeNet, Transcoder, and Avatar) using seven state-of-the-art LLMs. 
We report the percentage of successful translations for each source-target language pair, with success measured by functional equivalence.
Overall, GPT-4o~\cite{hurst2024gpto} demonstrates the strongest performance, achieving the highest average computation accuracy across all datasets: 83.33\% on CodeNet, 90.05\% on Transcoder, and 72.42\% on Avatar. This highlights its robust capability in handling diverse code translation tasks.
DeepSeek-Coder-V2~\cite{zhu2024deepseek} and Qwen 2.5-coder~\cite{yang2024qwen2.5} perform competitively, especially on CodeNet and Transcoder, with DeepSeek-Coder-V2 slightly outperforming Qwen 2.5-coder on average. For example, DeepSeek-Coder-V2 achieves 69.25\% on CodeNet and 76.14\% on Transcoder, compared to 63.17\% and 73.36\% respectively for Qwen 2.5-coder.

In contrast, models such as Llama 3.1~\cite{grattafiori2024llama}, Gemma 2~\cite{team2024gemma}, and Mistral~\cite{mistral7b} show lower computation accuracy, particularly on the Avatar dataset. Mistral, for instance, only achieves 18.90\% average on Avatar, indicating its limited ability to generalize to real-world, less-structured codebases.
A closer look at the direction-specific results reveals that translations involving Python as the target language tend to be more challenging. For instance, on CodeNet, the C++$\rightarrow$Python and Java$\rightarrow$Python tasks yield consistently lower computation accuracy across models. Conversely, translations between syntactically similar languages, such as C++ and Java, generally yield higher computation accuracy.
The Transcoder dataset yields overall higher computation accuracy, likely due to its more controlled and balanced nature, while the Avatar dataset proves the most difficult, reflecting the challenges of handling diverse and noisy code from real repositories.
These findings suggest that while recent LLMs show promising translation capabilities, there remains a performance gap between models and datasets. The poor performance is largely due to the hallucinations generated by the model during code translation. 
Therefore, in the next section, we conduct an in-depth quantitative and qualitative analysis of hallucination phenomena across different models, datasets, and language pairs, aiming to answer how hallucinations are distributed and which types of hallucinations dominate LLM-based code translation failures.

\begin{table*}[!t]
    \centering
    \small
    \caption{The performance of the Vanilla LLMs on code translation.}
    \adjustbox{width=\textwidth}{
    \begin{tabular}{lcccccccccc}
    \toprule
    \multirow{3}{*}{\textbf{Dataset}} & \multirow{3}{*}{\textbf{Source}} & \multirow{3}{*}{\textbf{Target}} & \multirow{3}{*}{\textbf{\#Samples}}  & \multicolumn{7}{c}{\textbf{\% Successful Translations}}  \\
    \cmidrule(l){5-11}
    & & & & \makecell{Qwen 2.5} & \makecell{DS-Coder} & \makecell{Llama 3.1} & \makecell{Gemma 2} & Mistral & Phi-4 & GPT-4o \\
    \midrule
    \multirow{6}{*}{CodeNet} 
    & C++ & Java & 200 & 76.00 & 85.50 & 58.50 & 63.00 & 39.00 & 78.50 & 87.50\\
    & C++ & Py & 200 & 43.50 & 57.50 & 38.50 & 49.00 & 27.50 & 63.50 & 74.00\\
    & Java & C++ & 200 & 68.50 & 74.00 & 48.50 & 60.00 & 37.50 & 72.50 & 89.50\\
    & Java & Py & 200 & 44.00 & 48.50 & 36.50 & 35.50 & 19.00 & 59.00 & 79.50\\
    & Py & C++ & 200 & 70.00 & 68.50 & 49.00 & 59.00 & 29.50 & 71.00 & 84.50\\
    & Py & Java & 200 & 77.00 & 81.50 & 57.00 & 59.00 & 49.00 & 75.00 & 85.00\\
    \midrule
    \rowcolor{gray!30} Avg (CodeNet) & - & - & - & 63.17 & 69.25 & 48.00 & 54.25 & 33.58 & 69.92 & 83.33 \\
    \midrule
    \multirow{6}{*}{Transcoder} 
    & C++ & Java & 482 & 82.99 & 70.33 & 70.95 & 45.64 & 58.30 & 65.35 & 87.55 \\
    & C++ & Py & 464 & 70.69 & 84.91 & 51.94 & 79.96 & 51.51 & 65.30 & 89.22 \\
    & Java & C++ & 467 & 81.37 & 75.37 & 83.73 & 84.80 & 61.67 & 71.09 & 95.29 \\
    & Java & Py & 464 & 75.22 & 86.64 & 69.18 & 83.19 & 54.53 & 70.04 & 91.59 \\
    & Py & C++ & 467 & 66.17 & 57.82 & 65.74 & 69.16 & 37.26 & 55.25 & 88.65 \\
    & Py & Java & 482 & 63.69 & 81.74 & 70.95 & 78.42 & 48.96 & 70.33 & 87.97 \\
    \midrule
    \rowcolor{gray!30} Avg (Transcoder) & - & - & - & 73.36 & 76.14 & 68.75 & 73.53 & 52.04 & 66.23 & 90.05
 \\
    \midrule
    \multirow{4}{*}{Avatar} 
    & Java & Py & 240 & 42.32 & 45.83 & 27.50 & 32.78 & 15.35 & 60.83 & 80.42\\
    & Java & C++ & 240 & 56.25 & 72.50 & 47.92 & 54.17 & 33.47 & 60.00 & 83.33\\
    & Py & Java & 239 & 50.63 & 53.14 & 32.22 & 27.20 & 15.90 & 56.90 & 59.83\\
    & Py & C++ & 239 & 41.42 & 43.51 & 22.59 & 27.62 & 10.88 & 46.03 & 66.11\\
    \midrule
     \rowcolor{gray!30} Avg (Avatar) & - & - & - & 47.66 &  53.75 & 32.56 & 35.44 & 18.90 & 55.94 & 72.42
  \\
    \bottomrule
    \end{tabular}
    }
    \label{comparision}
\end{table*}

\findingbox{
Significant performance disparities exist across LLMs in code translation, with code-specialized models and large closed-source models achieving substantially higher accuracy. Translation on language pairs with large syntactic discrepancies remain particularly challenging, such as Java-to-Python, indicating that severe syntactic and semantic differences further impede cross-language alignment.
}

\subsection{RQ2: How are hallucinations distributed across models and language pairs?}
As the core exploratory RQ targeting the research object, it is critical for revealing the intrinsic characteristics of code translation hallucinations. By quantifying the proportional distribution of hallucinations across the proposed taxonomy and its subcategories, this RQ identifies the dominant and high-frequency hallucination types, which is the prerequisite for designing targeted rather than one-size-fits-all mitigation strategies, and ensures the pertinence of subsequent hallucination alleviation research.

\begin{figure*}[t]
    \centering
    \begin{minipage}[t]{0.36\linewidth}
        \centering
        \includegraphics[width=\linewidth]{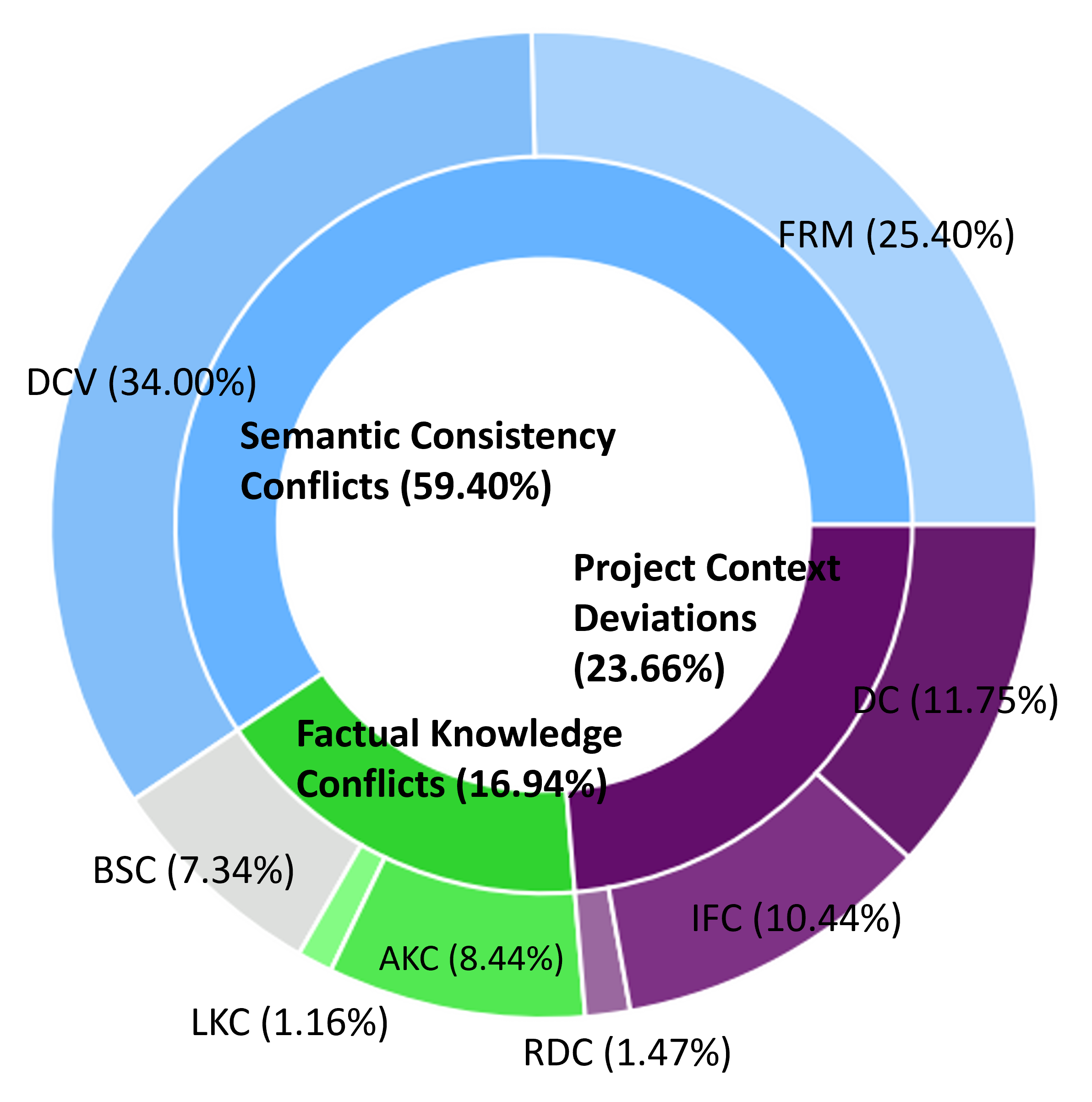}
        \caption{Hallucination Distribution
        }
        \label{fig:heat}
    \end{minipage}
    \hfill
    \begin{minipage}[t]{0.63\linewidth}
        \centering
        \includegraphics[width=\linewidth]{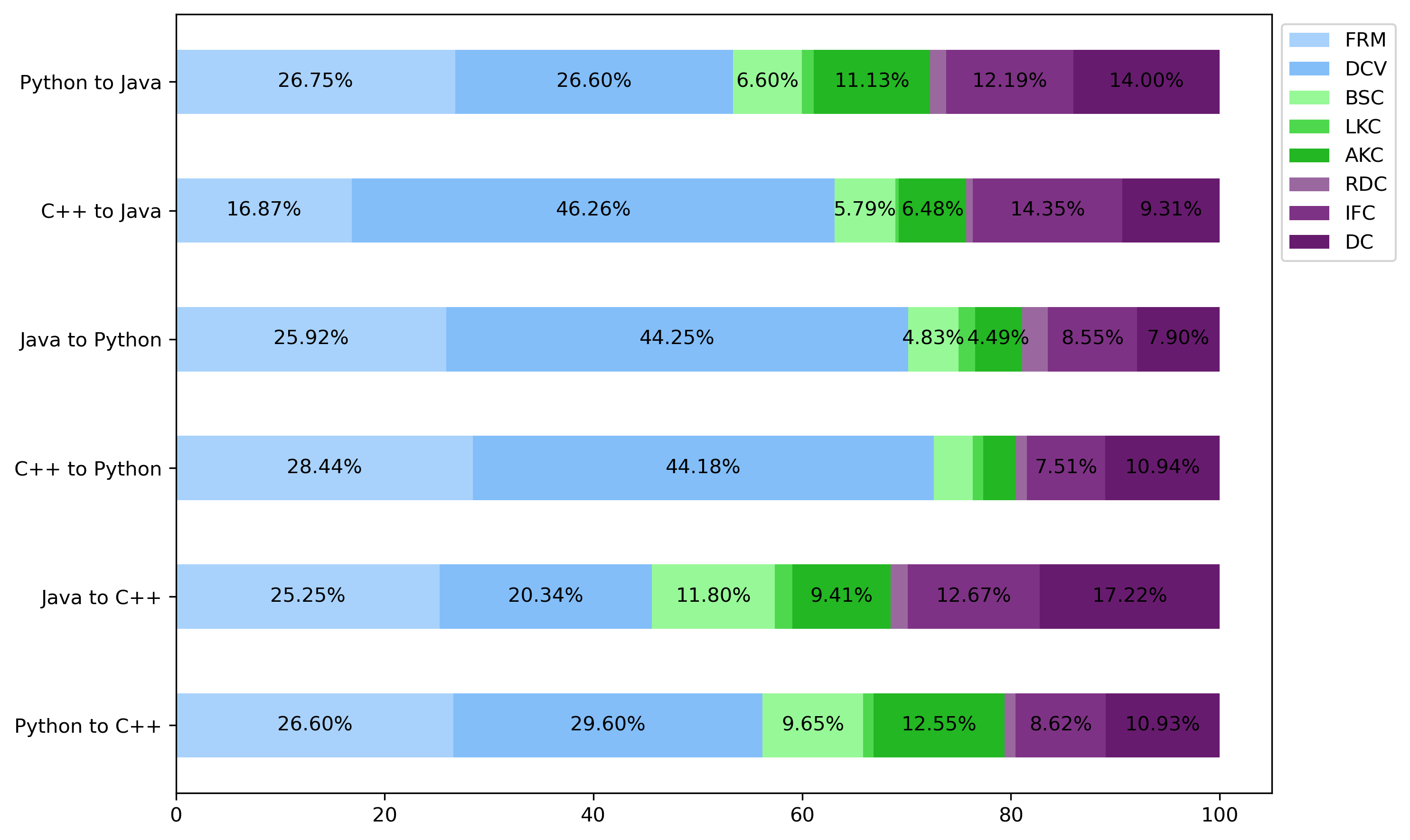}
        \caption{Distribution Across Programming Language Pairs}
        \label{fig:lang_dist}
    \end{minipage}
\end{figure*}

The distribution of hallucinations produced by the seven LLMs across the three code translation datasets is illustrated in Figure~\ref{fig:heat} and summarized in Table~\ref{tab:model_dataset_performance}. In the nested donut chart of Figure~\ref{fig:heat}, the inner ring represents broad categories of hallucinations, while the outer ring details specific subcategories, providing a granular view of error types in the models' outputs. Detailed numerical statistics for each category and model are presented in Table~\ref{tab:model_dataset_performance}.

Semantic Consistency Conflicts constitute the most dominant class of hallucinations, in which Functional Requirement Mismatch (FRM) and Data Consistency Violation (DCV) account for 25.41\% and 34.00\% of all errors, respectively. These hallucinations represent deeper semantic misunderstandings where the translated code either diverges from the intended functionality (FRM) or introduces inconsistencies in input-output behavior (DCV). 

In contrast, Factual Knowledge Conflicts, including Basic Syntax Conflicts (BSC), Library Knowledge Conflicts (LKC) and API Knowledge Conflicts (AKC), account for a smaller portion of hallucinations, collectively 16.94\%. Although less frequent than semantic hallucinations, factual hallucinations can still lead to functional failures, especially in contexts where precise API handling or exception management is critical. Their relatively lower prevalence suggests that LLMs have acquired a reasonable level of syntactic fluency and factual recall, likely due to exposure to large-scale programming corpora, but still fall short in edge cases and domain-specific conventions.

Finally, Project Context Deviations, including Identifier \& Function Confusion (IFC, 10.44\%), Dependency Conflicts (DC, 11.75\%), and Repetition \& Dead Code (RDC, 1.47\%), collectively account for nearly 30\% of the observed hallucinations, reflecting a notable challenge for LLMs in maintaining alignment with project-specific context. Although these hallucinations may not always lead to immediate failures, they undermine the reliability and maintainability of the generated code and may introduce subtle, hard-to-detect bugs.

\findingbox{
Semantic consistency conflicts account for the majority of hallucinations, indicating that preserving functional equivalence is far more challenging than maintaining syntactic correctness for LLMs. Data consistency violations and functional requirement mismatches are the two most frequent error patterns in code translation.
}

\begin{table}[htbp]
  \centering
  \caption{Average Hallucination Statistics Across Different Large Language Models}
  \resizebox{\linewidth}{!}{
    \begin{tabular}{llccccccccccc}
      \toprule
      model & dataset & 
      \cellcolor[HTML]{F0F7FF} FRM & 
      \cellcolor[HTML]{F0F7FF} DCV & 
      \cellcolor[HTML]{CCE5FF} SCC &
      \cellcolor[HTML]{FAFFFA} BSC & 
      \cellcolor[HTML]{FAFFFA} LKC & 
      \cellcolor[HTML]{FAFFFA} AKC & 
      \cellcolor[HTML]{F0F9F0} FKC &  %
      \cellcolor[HTML]{FAF5FF} RDC &   %
      \cellcolor[HTML]{FAF5FF} IFC &  %
      \cellcolor[HTML]{FAF5FF} DC &   %
      \cellcolor[HTML]{F0E5FF} PCD \\ %
      \midrule
      \multirow{3}{*}{Qwen 2.5} 
        & Avatar & \cellcolor[HTML]{F0F7FF} 33.52\% & \cellcolor[HTML]{F0F7FF} 17.85\% & \cellcolor[HTML]{CCE5FF} 51.37\% & \cellcolor[HTML]{FAFFFA} 8.01\% & \cellcolor[HTML]{FAFFFA} 0.55\% & \cellcolor[HTML]{FAFFFA} 10.20\% & \cellcolor[HTML]{F0F9F0} 18.76\% & \cellcolor[HTML]{FAF5FF} 2.55\% & \cellcolor[HTML]{FAF5FF} 12.39\% & \cellcolor[HTML]{FAF5FF} 14.94\% & \cellcolor[HTML]{F0E5FF} 29.87\% \\
        & CodeNet & \cellcolor[HTML]{F0F7FF} 21.41\% & \cellcolor[HTML]{F0F7FF} 32.85\% & \cellcolor[HTML]{CCE5FF} 54.26\% & \cellcolor[HTML]{FAFFFA} 6.44\% & \cellcolor[HTML]{FAFFFA} 0.21\% & \cellcolor[HTML]{FAFFFA} 11.43\% & \cellcolor[HTML]{F0F9F0} 18.09\% & \cellcolor[HTML]{FAF5FF} 0.62\% & \cellcolor[HTML]{FAF5FF} 12.27\% & \cellcolor[HTML]{FAF5FF} 14.76\% & \cellcolor[HTML]{F0E5FF} 27.65\% \\
        & Transcoder & \cellcolor[HTML]{F0F7FF} 19.08\% & \cellcolor[HTML]{F0F7FF} 62.61\% & \cellcolor[HTML]{CCE5FF} 81.69\% & \cellcolor[HTML]{FAFFFA} 2.43\% & \cellcolor[HTML]{FAFFFA} 0.26\% & \cellcolor[HTML]{FAFFFA} 3.46\% & \cellcolor[HTML]{F0F9F0} 6.15\% & \cellcolor[HTML]{FAF5FF} 0.38\% & \cellcolor[HTML]{FAF5FF} 6.40\% & \cellcolor[HTML]{FAF5FF} 5.38\% & \cellcolor[HTML]{F0E5FF} 12.16\% \\
      \midrule
      \multirow{3}{*}{DS-Coder} 
        & Avatar & \cellcolor[HTML]{F0F7FF} 36.81\% & \cellcolor[HTML]{F0F7FF} 15.54\% & \cellcolor[HTML]{CCE5FF} 52.35\% & \cellcolor[HTML]{FAFFFA} 9.00\% & \cellcolor[HTML]{FAFFFA} 1.84\% & \cellcolor[HTML]{FAFFFA} 7.98\% & \cellcolor[HTML]{F0F9F0} 18.81\% & \cellcolor[HTML]{FAF5FF} 4.29\% & \cellcolor[HTML]{FAF5FF} 12.88\% & \cellcolor[HTML]{FAF5FF} 11.66\% & \cellcolor[HTML]{F0E5FF} 28.83\% \\
        & CodeNet & \cellcolor[HTML]{F0F7FF} 29.22\% & \cellcolor[HTML]{F0F7FF} 24.94\% & \cellcolor[HTML]{CCE5FF} 54.16\% & \cellcolor[HTML]{FAFFFA} 11.84\% & \cellcolor[HTML]{FAFFFA} 2.02\% & \cellcolor[HTML]{FAFFFA} 5.79\% & \cellcolor[HTML]{F0F9F0} 19.65\% & \cellcolor[HTML]{FAF5FF} 1.76\% & \cellcolor[HTML]{FAF5FF} 11.84\% & \cellcolor[HTML]{FAF5FF} 12.59\% & \cellcolor[HTML]{F0E5FF} 26.20\% \\
        & Transcoder & \cellcolor[HTML]{F0F7FF} 25.72\% & \cellcolor[HTML]{F0F7FF} 60.49\% & \cellcolor[HTML]{CCE5FF} 86.21\% & \cellcolor[HTML]{FAFFFA} 1.44\% & \cellcolor[HTML]{FAFFFA} 0.72\% & \cellcolor[HTML]{FAFFFA} 2.01\% & \cellcolor[HTML]{F0F9F0} 4.17\% & \cellcolor[HTML]{FAF5FF} 1.15\% & \cellcolor[HTML]{FAF5FF} 3.45\% & \cellcolor[HTML]{FAF5FF} 5.03\% & \cellcolor[HTML]{F0E5FF} 9.63\% \\
      \midrule
      \multirow{3}{*}{Gemma 2}
        & Avatar & \cellcolor[HTML]{F0F7FF} 23.11\% & \cellcolor[HTML]{F0F7FF} 20.17\% & \cellcolor[HTML]{CCE5FF} 43.28\% & \cellcolor[HTML]{FAFFFA} 7.56\% & \cellcolor[HTML]{FAFFFA} 1.26\% & \cellcolor[HTML]{FAFFFA} 11.48\% & \cellcolor[HTML]{F0F9F0} 20.31\% & \cellcolor[HTML]{FAF5FF} 2.38\% & \cellcolor[HTML]{FAF5FF} 15.27\% & \cellcolor[HTML]{FAF5FF} 18.77\% & \cellcolor[HTML]{F0E5FF} 36.41\% \\
        & CodeNet & \cellcolor[HTML]{F0F7FF} 15.94\% & \cellcolor[HTML]{F0F7FF} 24.15\% & \cellcolor[HTML]{CCE5FF} 40.10\% & \cellcolor[HTML]{FAFFFA} 12.24\% & \cellcolor[HTML]{FAFFFA} 0.48\% & \cellcolor[HTML]{FAFFFA} 10.14\% & \cellcolor[HTML]{F0F9F0} 22.87\% & \cellcolor[HTML]{FAF5FF} 1.13\% & \cellcolor[HTML]{FAF5FF} 17.55\% & \cellcolor[HTML]{FAF5FF} 18.36\% & \cellcolor[HTML]{F0E5FF} 37.04\% \\
        & Transcoder & \cellcolor[HTML]{F0F7FF} 20.35\% & \cellcolor[HTML]{F0F7FF} 47.10\% & \cellcolor[HTML]{CCE5FF} 67.45\% & \cellcolor[HTML]{FAFFFA} 8.63\% & \cellcolor[HTML]{FAFFFA} 0.12\% & \cellcolor[HTML]{FAFFFA} 4.07\% & \cellcolor[HTML]{F0F9F0} 12.82\% & \cellcolor[HTML]{FAF5FF} 0.37\% & \cellcolor[HTML]{FAF5FF} 11.22\% & \cellcolor[HTML]{FAF5FF} 8.14\% & \cellcolor[HTML]{F0E5FF} 19.73\% \\
      \midrule
      \multirow{3}{*}{Llama 3.1}
        & Avatar & \cellcolor[HTML]{F0F7FF} 31.60\% & \cellcolor[HTML]{F0F7FF} 17.24\% & \cellcolor[HTML]{CCE5FF} 48.84\% & \cellcolor[HTML]{FAFFFA} 5.47\% & \cellcolor[HTML]{FAFFFA} 1.50\% & \cellcolor[HTML]{FAFFFA} 13.00\% & \cellcolor[HTML]{F0F9F0} 19.97\% & \cellcolor[HTML]{FAF5FF} 2.46\% & \cellcolor[HTML]{FAF5FF} 12.72\% & \cellcolor[HTML]{FAF5FF} 16.01\% & \cellcolor[HTML]{F0E5FF} 31.19\% \\
        & CodeNet & \cellcolor[HTML]{F0F7FF} 33.48\% & \cellcolor[HTML]{F0F7FF} 20.91\% & \cellcolor[HTML]{CCE5FF} 54.39\% & \cellcolor[HTML]{FAFFFA} 7.31\% & \cellcolor[HTML]{FAFFFA} 0.88\% & \cellcolor[HTML]{FAFFFA} 9.65\% & \cellcolor[HTML]{F0F9F0} 17.84\% & \cellcolor[HTML]{FAF5FF} 1.32\% & \cellcolor[HTML]{FAF5FF} 14.62\% & \cellcolor[HTML]{FAF5FF} 11.84\% & \cellcolor[HTML]{F0E5FF} 27.78\% \\
        & Transcoder & \cellcolor[HTML]{F0F7FF} 32.78\% & \cellcolor[HTML]{F0F7FF} 40.31\% & \cellcolor[HTML]{CCE5FF} 73.09\% & \cellcolor[HTML]{FAFFFA} 2.77\% & \cellcolor[HTML]{FAFFFA} 0.44\% & \cellcolor[HTML]{FAFFFA} 2.21\% & \cellcolor[HTML]{F0F9F0} 5.43\% & \cellcolor[HTML]{FAF5FF} 0.89\% & \cellcolor[HTML]{FAF5FF} 7.09\% & \cellcolor[HTML]{FAF5FF} 13.51\% & \cellcolor[HTML]{F0E5FF} 21.48\% \\
      \midrule
      \multirow{3}{*}{Mistral}
        & Avatar & \cellcolor[HTML]{F0F7FF} 22.78\% & \cellcolor[HTML]{F0F7FF} 14.33\% & \cellcolor[HTML]{CCE5FF} 37.11\% & \cellcolor[HTML]{FAFFFA} 9.90\% & \cellcolor[HTML]{FAFFFA} 3.92\% & \cellcolor[HTML]{FAFFFA} 17.11\% & \cellcolor[HTML]{F0F9F0} 30.93\% & \cellcolor[HTML]{FAF5FF} 2.89\% & \cellcolor[HTML]{FAF5FF} 12.16\% & \cellcolor[HTML]{FAF5FF} 16.91\% & \cellcolor[HTML]{F0E5FF} 31.96\% \\
        & CodeNet & \cellcolor[HTML]{F0F7FF} 22.50\% & \cellcolor[HTML]{F0F7FF} 16.15\% & \cellcolor[HTML]{CCE5FF} 38.65\% & \cellcolor[HTML]{FAFFFA} 8.38\% & \cellcolor[HTML]{FAFFFA} 3.43\% & \cellcolor[HTML]{FAFFFA} 17.66\% & \cellcolor[HTML]{F0F9F0} 29.47\% & \cellcolor[HTML]{FAF5FF} 0.30\% & \cellcolor[HTML]{FAF5FF} 15.94\% & \cellcolor[HTML]{FAF5FF} 15.64\% & \cellcolor[HTML]{F0E5FF} 31.89\% \\
        & Transcoder & \cellcolor[HTML]{F0F7FF} 24.74\% & \cellcolor[HTML]{F0F7FF} 46.63\% & \cellcolor[HTML]{CCE5FF} 71.37\% & \cellcolor[HTML]{FAFFFA} 1.57\% & \cellcolor[HTML]{FAFFFA} 1.23\% & \cellcolor[HTML]{FAFFFA} 7.02\% & \cellcolor[HTML]{F0F9F0} 9.82\% & \cellcolor[HTML]{FAF5FF} 0.95\% & \cellcolor[HTML]{FAF5FF} 6.68\% & \cellcolor[HTML]{FAF5FF} 11.18\% & \cellcolor[HTML]{F0E5FF} 18.81\% \\
        \midrule
      \multirow{3}{*}{Phi-4} 
        & Avatar & \cellcolor[HTML]{F0F7FF} 26.72\% & \cellcolor[HTML]{F0F7FF} 15.73\% & \cellcolor[HTML]{CCE5FF} 42.46\% & \cellcolor[HTML]{FAFFFA} 26.94\% & \cellcolor[HTML]{FAFFFA} 0.22\% & \cellcolor[HTML]{FAFFFA} 8.41\% & \cellcolor[HTML]{F0F9F0} 35.56\% & \cellcolor[HTML]{FAF5FF} 3.23\% & \cellcolor[HTML]{FAF5FF} 10.56\% & \cellcolor[HTML]{FAF5FF} 8.19\% & \cellcolor[HTML]{F0E5FF} 21.98\% \\
        & CodeNet & \cellcolor[HTML]{F0F7FF} 19.44\% & \cellcolor[HTML]{F0F7FF} 25.58\% & \cellcolor[HTML]{CCE5FF} 45.01\% & \cellcolor[HTML]{FAFFFA} 25.32\% & \cellcolor[HTML]{FAFFFA} 0.77\% & \cellcolor[HTML]{FAFFFA} 9.46\% & \cellcolor[HTML]{F0F9F0} 35.55\% & \cellcolor[HTML]{FAF5FF} 0.77\% & \cellcolor[HTML]{FAF5FF} 8.18\% & \cellcolor[HTML]{FAF5FF} 10.49\% & \cellcolor[HTML]{F0E5FF} 19.44\% \\
        & Transcoder & \cellcolor[HTML]{F0F7FF} 14.34\% & \cellcolor[HTML]{F0F7FF} 74.69\% & \cellcolor[HTML]{CCE5FF} 89.04\% & \cellcolor[HTML]{FAFFFA} 1.02\% & \cellcolor[HTML]{FAFFFA} 0.00\% & \cellcolor[HTML]{FAFFFA} 2.05\% & \cellcolor[HTML]{F0F9F0} 3.07\% & \cellcolor[HTML]{FAF5FF} 0.41\% & \cellcolor[HTML]{FAF5FF} 3.69\% & \cellcolor[HTML]{FAF5FF} 3.79\% & \cellcolor[HTML]{F0E5FF} 7.89\% \\
      \midrule
      \multirow{3}{*}{GPT-4o}
        & Avatar & \cellcolor[HTML]{F0F7FF} 37.50\% & \cellcolor[HTML]{F0F7FF} 10.42\% & \cellcolor[HTML]{CCE5FF} 47.92\% & \cellcolor[HTML]{FAFFFA} 19.79\% & \cellcolor[HTML]{FAFFFA} 0.35\% & \cellcolor[HTML]{FAFFFA} 8.33\% & \cellcolor[HTML]{F0F9F0} 28.47\% & \cellcolor[HTML]{FAF5FF} 4.51\% & \cellcolor[HTML]{FAF5FF} 10.07\% & \cellcolor[HTML]{FAF5FF} 9.03\% & \cellcolor[HTML]{F0E5FF} 23.61\% \\
        & CodeNet & \cellcolor[HTML]{F0F7FF} 26.73\% & \cellcolor[HTML]{F0F7FF} 22.58\% & \cellcolor[HTML]{CCE5FF} 49.31\% & \cellcolor[HTML]{FAFFFA} 5.99\% & \cellcolor[HTML]{FAFFFA} 1.84\% & \cellcolor[HTML]{FAFFFA} 13.36\% & \cellcolor[HTML]{F0F9F0} 21.20\% & \cellcolor[HTML]{FAF5FF} 1.38\% & \cellcolor[HTML]{FAF5FF} 16.59\% & \cellcolor[HTML]{FAF5FF} 11.52\% & \cellcolor[HTML]{F0E5FF} 29.49\% \\
        & Transcoder & \cellcolor[HTML]{F0F7FF} 42.62\% & \cellcolor[HTML]{F0F7FF} 39.60\% & \cellcolor[HTML]{CCE5FF} 82.21\% & \cellcolor[HTML]{FAFFFA} 2.01\% & \cellcolor[HTML]{FAFFFA} 0.34\% & \cellcolor[HTML]{FAFFFA} 3.02\% & \cellcolor[HTML]{F0F9F0} 5.37\% & \cellcolor[HTML]{FAF5FF} 1.01\% & \cellcolor[HTML]{FAF5FF} 6.71\% & \cellcolor[HTML]{FAF5FF} 4.70\% & \cellcolor[HTML]{F0E5FF} 12.42\% \\
      \bottomrule
    \end{tabular}
  }
  \label{tab:model_dataset_performance}
\end{table}

\subsection{RQ3: What is the impact of model scale and architecture on hallucinations in code translation?}
This RQ fills the research gap of lacking systematic analysis on the correlation between LLM attributes and code translation hallucinations. It is essential for disentangling how model scale and architectural design affect hallucination rates and type preferences, and its findings provide practical guidance for model selection in real-world code translation applications and valuable insights for optimizing LLM architectures to reduce code-related hallucinations.

Table~\ref{tab:model_dataset_performance} presents the hallucination distribution of the seven evaluated LLMs across datasets.
Overall, Mistral consistently exhibits the highest hallucination counts across most translation directions, with a relatively high rate of API Knowledge Conflicts (AKC), suggesting a tendency to hallucinate or misuse APIs even when general syntax and structure are preserved.
In contrast, GPT-4o exhibits the lowest overall hallucination rate, demonstrating strong reliability across translation pairs. However, its performance is primarily constrained by Semantic Consistency Conflicts, indicating that while GPT-4o excels in factual correctness and context consistency, it occasionally struggles to preserve the intended behavior or semantics of the source code.
DeepSeek-Coder ranks second only to GPT-4o in minimizing hallucinations, demonstrating strong generalization across a range of translation tasks. It is closely followed by Qwen 2.5-coder and Phi 4, both of which exhibit relatively stable performance, though with slightly greater fluctuations across hallucination types, particularly in categories related to project context and factual knowledge.
Gemma 2 exhibits relatively strong performance on the C++ to Python translation task, with a noticeably lower number of hallucinations compared to other language pairs. This indicates that Gemma 2 is particularly well-suited for conversions between C++ and Python, potentially due to more effective pattern alignment or training exposure in these languages. While LLaMA 3.1 shows weaker performance overall, it achieves competitive results in the C++ to Java translation task, suggesting that its strengths may be more pronounced in specific language pairs or syntactic structures.

\findingbox{
Model family, scale, and code-oriented pre-training are associated with different hallucination profiles, yet even the strongest evaluated models still produce semantic errors. This suggests that preserving deep program logic remains a major challenge for current LLM-based code translation.
}

\subsection{RQ4: What is the impact of type of programming languages on hallucinations in code translation?}
This RQ is pivotal for exploring the scenario-specific characteristics of code translation hallucinations. It analyzes how programming language properties (e.g., static/dynamic typing) and language pairings shape hallucination patterns, and identifies high-risk translation pairs and their typical hallucination subtypes. The results can guide practitioners to predict and avoid hallucination risks in specific translation scenarios, and also provide a basis for tailoring mitigation strategies for different language pairs.

To better understand how hallucinations vary with different programming languages, we analyze the distribution of hallucination types across six translation directions, as shown in Figure~\ref{fig:lang_dist}. The results indicate that hallucination patterns are not uniformly distributed but are influenced by both the source and target languages.

Among all translation directions, hallucinations in the Java to Python, C++ to Python, and C++ to Java tasks are dominated by Data Consistency Violations (DCV), accounting for 44.25\%, 44.18\% and 46.26\%, respectively. These results suggest that LLMs often struggle to preserve input-output behavior and state consistency when translating from statically typed languages into dynamically typed languages. In the case of C++ to Java, although both languages are statically typed, translation errors may stem from differences in how input and output are handled across programming languages, resulting in divergent behavior.

When translating Python to Java and Python to C++, the proportion of hallucinations related to API Knowledge Conflicts (AKC) increases. Specifically, API conflicts account for 12.55\% in Python to C++ and 11.13\% in Python to Java. This suggests that LLMs face substantial challenges in accurately resolving and translating the flexible, dynamically typed constructs of Python into the more explicit, rigidly defined APIs and dependency structures typical of statically typed target languages.

In the Java to C++ and Python to C++ translations, hallucinations are more evenly distributed across various categories. This implies that while translation to C++ introduces several hallucinations, no single type of hallucination overwhelmingly dominates, likely due to the complexity and strict requirements of the language imposing a wide range of potential failure points.

\findingbox{
The source and target programming languages influence hallucination patterns in code translation. Translations from statically typed languages (Java, C++) to dynamically typed languages (Python) are mainly affected by Data Consistency Violations (DCV), while translations from dynamically typed languages to statically typed languages exhibit more API Knowledge Conflicts (AKC), reflecting challenges in mapping flexible constructs to rigid target-language APIs.
}

\section{Hallucination Mitigation}
This section addresses the critical gap in existing literature regarding code translation hallucination mitigation frameworks and serves as the primary application-oriented inquiry of this study. Synthesizing our previous analysis of hallucination taxonomies and the limitations of static methodologies, we investigate the inherent reflection capabilities of LLMs. Specifically,  we propose a new approach termed HaluCodeTran, which leverages retrieval-augmentation to supplement LLMs with relevant code translation knowledge and a reflective iterative refinement strategy to identify and correct errors. 
In the following sections, we describe the technical details of baseline methods and HaluCodeTran and the quantitative experiments on hallucination mitigation.

\subsection{HaluCodeTran}

HaluCodeTran integrates retrieval-augmented translation with a feedback loop that prompts the model to examine its intermediate outputs against the source logic and the hallucination taxonomy. 
The process comprises three primary phases: 1) \textit{Knowledge Augmentation}, 2) \textit{Code Translation}, and 3) \textit{Recursive Self-Reflection}.

\textbf{1) Knowledge Augmentation via Contextual Grounding. }
The execution begins with the acquisition of the input source code $x$. To mitigate potential hallucinations arising from cross-language mapping ambiguities, we first perform knowledge augmentation. Specifically, we leverage a demonstrated knowledge base $\mathcal{D}$, containing high-quality pairs of source and target code. Following established practices~\cite{zhao2023felm}, we adopt the BM25 algorithm~\cite{bm25} to retrieve a subset of relevant examples $\mathcal{S}$ based on their lexical and structural relevance scores to $x$. These examples $\mathcal{S}$ act as a repository of \textit{implicit knowledge}, encompassing target language syntax, library usage, and idiomatic conversion patterns. By providing these grounded references, we effectively narrow the model's search space and provide a pedagogical basis for the subsequent translation.

\textbf{2) Initial Code Translation. }
Building upon the augmented context, the Code Translation phase aims to synthesize a functionally equivalent version of $x$. By conditioning the model on both the original source code $x$ and the retrieved knowledge subset $\mathcal{S}$, the translator generates an initial translation candidate $\hat{y}^0$:
\begin{equation}
    \hat{y}^0 = \text{Translate}(x, \mathcal{S})
\end{equation}
While $\mathcal{S}$ provides a strong prior, the inherent intricacies of code semantics and latent logic ambiguities mean that $\hat{y}^0$ may still harbor semantic inconsistencies, factual errors, or project-specific deviations as identified in our taxonomy.

\textbf{3) Recursive Self-Reflection and Refinement.}
To systematically rectify these hallucinations, HaluCodeTran enters a recursive self-reflection loop. In each iteration $i$, the model receives four inputs: the source program $x$, the current translation $\hat{y}^i$, the retrieved examples $\mathcal{S}$, and the taxonomy definitions in Section~\ref{scc}--\ref{pcd}. It then produces a reflection report $\mathcal{R}^i$ that checks whether $\hat{y}^i$ preserves source-program behavior, uses valid target-language syntax and APIs, and avoids project-context deviations. Unlike an execution-based repair loop, this reflection step does not observe hidden test results; it relies only on the source code, retrieved examples, and taxonomy-guided inspection.
If the reflection report does not identify any taxonomy category, $\text{IsConsistent}(\mathcal{R}^i)$ returns true and the process terminates. Otherwise, the model distills a structured critique $\mathcal{T}^i$ from the report. Each critique item follows the format \lstinline|<category: location: description: repair hint>|, where \lstinline|category| is one of the hallucination subtypes, \lstinline|location| identifies the suspicious target-code fragment, \lstinline|description| explains the mismatch, and \lstinline|repair hint| states the required source-aligned revision. This diagnostic feedback directly guides the subsequent refinement step:
\begin{equation}
    \hat{y}^{i+1} = \text{Refine}(\hat{y}^i, \mathcal{T}^i, x)
\end{equation}
This cycle of reflection, critique, and refinement iterates until the reflection step reports no detected hallucination or the maximum step $M$ is reached. The same $M$, retrieval size, and prompting budget are used across models in the mitigation study. The pseudo-code of HaluCodeTran is shown in Algorithm~\ref{pseduocode}.

\begin{algorithm}[h]
\caption{The pseudo-code of HaluCodeTran with Reflective Refinement}
\begin{algorithmic}[1]
\REQUIRE Input code $x$, Demonstrated Knowledge Base $\mathcal{D}$, maximum steps $M$
\ENSURE Refined translated code $y$
\STATE $\mathcal{S} \gets \text{RetrieveRelevantExamples}(x, \mathcal{D})$ \COMMENT{Contextual grounding via BM25}
\STATE $\hat{y}^0 \gets \text{Translate}(x, \mathcal{S})$ \COMMENT{Initial translation candidate}
\FOR{$i = 0$ \textbf{to} $M-1$}
    \STATE $\mathcal{R}^i \gets \text{Reflect}(\hat{y}^i, x, \mathcal{S})$ \COMMENT{Self-reflection for hallucinations}
    \IF{$\text{IsConsistent}(\mathcal{R}^i)$} 
        \RETURN $\hat{y}^i$ \COMMENT{Termination on consistency}
    \ELSE
        \STATE $\mathcal{T}^i \gets \text{ExtractCritique}(\mathcal{R}^i)$ \COMMENT{Knowledge distillation from reflection}
        \STATE $\hat{y}^{i+1} \gets \text{Refine}(\hat{y}^i, \mathcal{T}^i, x)$ \COMMENT{Iterative reflective refinement}
    \ENDIF
\ENDFOR
\RETURN $\hat{y}^M$ \COMMENT{Final refined output}
\end{algorithmic}
\label{pseduocode}
\end{algorithm}

\subsection{Baseline Methods of Hallucination Mitigation}

We perform a series of experiments to validate whether we can guide LLMs to alleviate hallucinations via prompt engineering.

\begin{itemize}
    \item Vanilla: is the standalone LLM without any additional preprocessing modules or external retrievers. Vanilla LLMs represent the original capabilities of LLMs.
    \item Chain-of-Thought(CoT)~\cite{wei2022chain}: is a prompting technique to encourage multi-step reasoning by explicitly guiding the LLMs to "think" step-by-step before answering.
    \item Chain-of-Verification(CoVe)~\cite{chainofverification}: prompts LLMs to draft the initial response, plan verification questions, answer the question, and generate the final verified response.
    \item Retrieval-Augmentation(RA): following the widely-used setting~\cite{zhao2023felm}, uses LLMs with the BM25 algorithm~\cite{bm25} to retrieve the most relevant texts as input to vanilla LLMs. 
\end{itemize}

We evaluate hallucination mitigation on CodeNet because it provides balanced language-pair coverage and executable test cases for all six translation directions. For each evaluated model, we construct a retrieval database from source programs and their successfully translated target programs in the model's successful pool. The evaluation set contains the remaining source programs that were not successfully translated by the same model under the vanilla setting. During retrieval, the query program itself is excluded from the candidate pool, and the model receives only retrieved source-target examples rather than test outputs or reference explanations. This setting evaluates whether a mitigation method can repair cases that are difficult for the original model while avoiding direct test-feedback repair.
For the Retrieval-Augmentation baseline, the 5-shot setting is chosen based on its superior performance in preliminary experiments. To ensure fairness, we also apply the same 5-shot retrieved context in the initial translation phase of HaluCodeTran. CoT, CoVe, RA, and HaluCodeTran are evaluated on the same per-model test split, using the same base model and the same execution-based CA metric.

\subsection{Evaluation Results of Hallucination Mitigation}

We compare HaluCodeTran with representative prompt-based and retrieval-based mitigation methods, including CoT, CoVe, and RA, across models and language pairs. This comparison is intended to test whether taxonomy-aware reflective refinement adds value beyond generic reasoning prompts, generic verification prompts, and retrieval alone.

\begin{table*}[t]
    \centering
    \caption{Performance of different models and mitigation techniques on CodeNet with improvements relative to Vanilla in terms of Computational Accuracy (darkgreen brackets indicate absolute percentage-point gains).}
    \label{tab:mitigation-performance}
    \adjustbox{width=\textwidth}{%
    \begin{tabular}{lllllllll}
        \toprule
        \textbf{Model} & \textbf{Technique} & \textbf{C++ to Java} & \textbf{Java to C++} & \textbf{C++ to Py} & \textbf{Py to C++} & \textbf{Java to Py} & \textbf{Py to Java} & \textbf{Average} \\
        \midrule
        \multirow{5}{*}{Gemma 2}
        & Vanilla & 63.00 & 60.00 & 49.00 & 59.00 & 35.50 & 59.00 & 54.25 \\
        & CoT & 71.00$_{\textcolor{darkgreen}{(+8.00)}}$ & 70.50$_{\textcolor{darkgreen}{(+10.50)}}$ & 56.00$_{\textcolor{darkgreen}{(+7.00)}}$ & 68.00$_{\textcolor{darkgreen}{(+9.00)}}$ & 39.50$_{\textcolor{darkgreen}{(+4.00)}}$ & \textbf{72.50}$_{\textcolor{darkgreen}{(+13.50)}}$ & 62.92$_{\textcolor{darkgreen}{(+8.67)}}$ \\
        & CoVe & 68.50$_{\textcolor{darkgreen}{(+5.50)}}$ & 69.50$_{\textcolor{darkgreen}{(+9.50)}}$ & 56.00$_{\textcolor{darkgreen}{(+7.00)}}$ & 66.00$_{\textcolor{darkgreen}{(+7.00)}}$ & 43.50$_{\textcolor{darkgreen}{(+8.00)}}$ & 66.00$_{\textcolor{darkgreen}{(+7.00)}}$ & 61.58$_{\textcolor{darkgreen}{(+7.33)}}$ \\
        & RA & 70.00$_{\textcolor{darkgreen}{(+7.00)}}$ & 71.00$_{\textcolor{darkgreen}{(+11.00)}}$ & 55.50$_{\textcolor{darkgreen}{(+6.50)}}$ & 65.50$_{\textcolor{darkgreen}{(+6.50)}}$ & 39.00$_{\textcolor{darkgreen}{(+3.50)}}$ & \textbf{72.50}$_{\textcolor{darkgreen}{(+13.50)}}$ & 62.25$_{\textcolor{darkgreen}{(+8.00)}}$ \\
        \cline{2-9}
        & HaluCodeTran & \textbf{71.50}$_{\textcolor{darkgreen}{(+8.50)}}$ & \textbf{72.50}$_{\textcolor{darkgreen}{(+12.50)}}$ & \textbf{58.00}$_{\textcolor{darkgreen}{(+9.00)}}$ & \textbf{69.00}$_{\textcolor{darkgreen}{(+10.00)}}$ & \textbf{45.50}$_{\textcolor{darkgreen}{(+10.00)}}$ & \textbf{72.50}$_{\textcolor{darkgreen}{(+13.50)}}$ & \textbf{64.83}$_{\textcolor{darkgreen}{(+10.58)}}$ \\
        \midrule
        \multirow{5}{*}{Llama 3.1}
        & Vanilla & 58.50 & 48.50 & 38.50 & 49.00 & 36.50 & 57.00 & 48.00 \\
        & CoT & 68.50$_{\textcolor{darkgreen}{(+10.00)}}$ & 58.50$_{\textcolor{darkgreen}{(+10.00)}}$ & 47.00$_{\textcolor{darkgreen}{(+8.50)}}$ & 58.00$_{\textcolor{darkgreen}{(+9.00)}}$ & 45.00$_{\textcolor{darkgreen}{(+8.50)}}$ & 65.00$_{\textcolor{darkgreen}{(+8.00)}}$ & 57.00$_{\textcolor{darkgreen}{(+9.00)}}$ \\
        & CoVe & 66.00$_{\textcolor{darkgreen}{(+7.50)}}$ & 59.50$_{\textcolor{darkgreen}{(+11.00)}}$ & 54.50$_{\textcolor{darkgreen}{(+16.00)}}$ & 59.00$_{\textcolor{darkgreen}{(+10.00)}}$ & 44.50$_{\textcolor{darkgreen}{(+8.00)}}$ & 67.00$_{\textcolor{darkgreen}{(+10.00)}}$ & 58.42$_{\textcolor{darkgreen}{(+10.42)}}$ \\
        & RA & 62.00$_{\textcolor{darkgreen}{(+3.50)}}$ & 59.50$_{\textcolor{darkgreen}{(+11.00)}}$ & 54.00$_{\textcolor{darkgreen}{(+15.50)}}$ & 52.00$_{\textcolor{darkgreen}{(+3.00)}}$ & 45.00$_{\textcolor{darkgreen}{(+8.50)}}$ & 68.00$_{\textcolor{darkgreen}{(+11.00)}}$ & 56.75$_{\textcolor{darkgreen}{(+8.75)}}$ \\
        \cline{2-9}
        & HaluCodeTran & \textbf{69.00}$_{\textcolor{darkgreen}{(+10.50)}}$ & \textbf{62.00}$_{\textcolor{darkgreen}{(+13.50)}}$ & \textbf{55.00}$_{\textcolor{darkgreen}{(+16.50)}}$ & \textbf{60.50}$_{\textcolor{darkgreen}{(+11.50)}}$ & \textbf{50.50}$_{\textcolor{darkgreen}{(+14.00)}}$ & \textbf{69.50}$_{\textcolor{darkgreen}{(+12.50)}}$ & \textbf{61.08}$_{\textcolor{darkgreen}{(+13.08)}}$ \\
        \midrule
        \multirow{5}{*}{Mistral}
        & Vanilla & 39.00 & 37.50 & 27.50 & 29.50 & 19.00 & 49.00 & 33.58 \\
        & CoT & 46.50$_{\textcolor{darkgreen}{(+7.50)}}$ & 42.00$_{\textcolor{darkgreen}{(+4.50)}}$ & 36.50$_{\textcolor{darkgreen}{(+9.00)}}$ & 42.50$_{\textcolor{darkgreen}{(+13.00)}}$ & 26.00$_{\textcolor{darkgreen}{(+7.00)}}$ & 57.50$_{\textcolor{darkgreen}{(+8.50)}}$ & 41.83$_{\textcolor{darkgreen}{(+8.25)}}$ \\
        & CoVe & 51.00$_{\textcolor{darkgreen}{(+12.00)}}$ & 49.50$_{\textcolor{darkgreen}{(+12.00)}}$ & 42.00$_{\textcolor{darkgreen}{(+14.50)}}$ & 36.00$_{\textcolor{darkgreen}{(+6.50)}}$ & 31.00$_{\textcolor{darkgreen}{(+12.00)}}$ & 57.00$_{\textcolor{darkgreen}{(+8.00)}}$ & 44.42$_{\textcolor{darkgreen}{(+10.83)}}$ \\
        & RA & 59.00$_{\textcolor{darkgreen}{(+20.00)}}$ & 56.50$_{\textcolor{darkgreen}{(+19.00)}}$ & 35.50$_{\textcolor{darkgreen}{(+8.00)}}$ & 51.50$_{\textcolor{darkgreen}{(+22.00)}}$ & 30.00$_{\textcolor{darkgreen}{(+11.00)}}$ & 56.00$_{\textcolor{darkgreen}{(+7.00)}}$ & 48.08$_{\textcolor{darkgreen}{(+14.50)}}$ \\
        \cline{2-9}
        & HaluCodeTran & \textbf{59.50}$_{\textcolor{darkgreen}{(+20.50)}}$ & \textbf{57.00}$_{\textcolor{darkgreen}{(+19.50)}}$ & \textbf{43.00}$_{\textcolor{darkgreen}{(+15.50)}}$ & \textbf{52.50}$_{\textcolor{darkgreen}{(+23.00)}}$ & \textbf{32.00}$_{\textcolor{darkgreen}{(+13.00)}}$ & \textbf{58.00}$_{\textcolor{darkgreen}{(+9.00)}}$ & \textbf{50.33}$_{\textcolor{darkgreen}{(+16.75)}}$ \\
        \bottomrule
    \end{tabular}%
    }
\end{table*}

Table~\ref{tab:mitigation-performance} presents the code translation performance of different hallucination mitigation techniques across three models (Llama 3.1, Gemma 2, Mistral) on the CodeNet dataset. 
Regarding the experimental results, three key findings can be summarized as follows. First, all mitigation methods improve average CA over vanilla translation for the three evaluated models. Mistral, which has the weakest vanilla performance on CodeNet, improves from 33.58\% to 50.33\% under HaluCodeTran. Llama 3.1 improves from 48.00\% to 61.08\%, and Gemma 2 improves from 54.25\% to 64.83\%. These results indicate that contextual prompting and refinement can partially compensate for weak base translation capability, especially for smaller general-purpose models.

Second, the three baseline mitigation strategies show different strengths. CoT is the strongest non-HaluCodeTran baseline for Gemma 2 on average, improving CA by 8.67 percentage points over vanilla and matching the best result on Python$\rightarrow$Java. CoVe is the strongest non-HaluCodeTran baseline for Llama 3.1, with a 10.42 percentage-point average gain and the largest baseline gain on C++$\rightarrow$Python. RA is particularly effective for Mistral, improving average CA by 14.50 percentage points and achieving a 22.00 percentage-point gain on Python$\rightarrow$C++. These differences show that reasoning, verification, and retrieval address different parts of the code-translation failure surface, but no single generic baseline dominates across all models and language pairs.

Third, HaluCodeTran achieves the highest average CA for all three evaluated models. Compared with RA, it adds 2.58 percentage points on Gemma 2, 4.33 points on Llama 3.1, and 2.25 points on Mistral. This suggests that retrieval provides useful grounding, while taxonomy-aware reflection further helps the model localize and revise semantic inconsistencies, factual errors, and context deviations. The gains are not uniform across all language pairs, so we interpret HaluCodeTran as a practical mitigation framework rather than a complete solution to code-translation hallucination.

\findingbox{
Contextual prompting can mitigate hallucinations in code translation, but its efficacy is limited. CoT specializes in translation logic standardization, CoVe in error correction, and RA in knowledge supplementation, providing clear references for subsequent hallucination mitigation framework design. LLMs can effectively rectify translation hallucinations through guided reflection mechanisms. HaluCodeTran achieves superior and stable improvements by combining retrieval augmentation and iterative self-refinement, validating that taxonomy-aware feedback strongly facilitates autonomous error correction.
}

\begin{table*}[h]
    \centering
    \footnotesize
    \caption{The ability of different methods for hallucination mitigation on the CodeNet dataset in terms of the number of hallucination samples.}
    \label{tab:mitigation-transport}
    \adjustbox{width=\textwidth}{
    \begin{tabular}{llll|lll|lll|l}
        \toprule
        \multirow{2}{*}{\textbf{Model}} & \multirow{2}{*}{\textbf{Baseline}} & \multicolumn{8}{c}{\textbf{\# Hallucination}}\\
        \cmidrule(r){3-11}
         & & FRM & DCV & BSC & LKC & AKC & RDC & IFC & DC & Total\\
        \midrule 
        \multirow{5}{*}{Gemma 2} & Vanilla & 
        99 & 150 & 76 & 3 &	63 & 7 & 109 & 114 & 621\\
        & CoT &
        88$_{\textcolor{darkgreen}{(\ensuremath{\downarrow}11\%)}}$ & 116$_{\textcolor{darkgreen}{(\ensuremath{\downarrow}23\%)}}$ & 74$_{\textcolor{darkgreen}{(\ensuremath{\downarrow}3\%)}}$ & 2$_{\textcolor{darkgreen}{(\ensuremath{\downarrow}33\%)}}$ & 53$_{\textcolor{darkgreen}{(\ensuremath{\downarrow}16\%)}}$ &	8$_{\textcolor{red}{(\ensuremath{\uparrow}14\%)}}$ & 93$_{\textcolor{darkgreen}{(\ensuremath{\downarrow}15\%)}}$ & 100$_{\textcolor{darkgreen}{(\ensuremath{\downarrow}12\%)}}$ & 534$_{\textcolor{darkgreen}{(\ensuremath{\downarrow}14\%)}}$\\
        & CoVe & 
        72$_{\textcolor{darkgreen}{(\ensuremath{\downarrow}27\%)}}$ & 132$_{\textcolor{darkgreen}{(\ensuremath{\downarrow}12\%)}}$ & 73$_{\textcolor{darkgreen}{(\ensuremath{\downarrow}4\%)}}$ &	3$_{\textcolor{gray}{(0\%)}}$ &	65$_{\textcolor{red}{(\ensuremath{\uparrow}3\%)}}$ & 2$_{\textcolor{darkgreen}{(\ensuremath{\downarrow}71\%)}}$ & 78$_{\textcolor{darkgreen}{(\ensuremath{\downarrow}28\%)}}$ & 83$_{\textcolor{darkgreen}{(\ensuremath{\downarrow}27\%)}}$ & 508$_{\textcolor{darkgreen}{(\ensuremath{\downarrow}18\%)}}$\\
        & RA & 
        85$_{\textcolor{darkgreen}{(\ensuremath{\downarrow}14\%)}}$ & 123$_{\textcolor{darkgreen}{(\ensuremath{\downarrow}18\%)}}$ & 75$_{\textcolor{darkgreen}{(\ensuremath{\downarrow}1\%)}}$ &	0$_{\textcolor{darkgreen}{(\ensuremath{\downarrow}100\%)}}$ &	72$_{\textcolor{red}{(\ensuremath{\uparrow}14\%)}}$ & 3$_{\textcolor{darkgreen}{(\ensuremath{\downarrow}57\%)}}$ & 77$_{\textcolor{darkgreen}{(\ensuremath{\downarrow}29\%)}}$ & 95$_{\textcolor{darkgreen}{(\ensuremath{\downarrow}17\%)}}$ & 530$_{\textcolor{darkgreen}{(\ensuremath{\downarrow}15\%)}}$\\
        \cline{2-11}
        & \textbf{HaluCodeTran} &
        79$_{\textcolor{darkgreen}{(\ensuremath{\downarrow}20\%)}}$ & 111$_{\textcolor{darkgreen}{(\ensuremath{\downarrow}26\%)}}$ & 43$_{\textcolor{darkgreen}{(\ensuremath{\downarrow}43\%)}}$ &	0$_{\textcolor{darkgreen}{(\ensuremath{\downarrow}100\%)}}$ & 56$_{\textcolor{darkgreen}{(\ensuremath{\downarrow}11\%)}}$ & 4$_{\textcolor{darkgreen}{(\ensuremath{\downarrow}43\%)}}$ & 56$_{\textcolor{darkgreen}{(\ensuremath{\downarrow}49\%)}}$ & 72$_{\textcolor{darkgreen}{(\ensuremath{\downarrow}37\%)}}$ & 421$_{\textcolor{darkgreen}{(\ensuremath{\downarrow}32\%)}}$\\
        \midrule
        \multirow{5}{*}{Llama 3.1} & Vanilla & 
        229	& 143 & 50 & 6 & 66 & 9 & 100 &	81 & 684\\
        & CoT & 
        174$_{\textcolor{darkgreen}{(\ensuremath{\downarrow}24\%)}}$ & 133$_{\textcolor{darkgreen}{(\ensuremath{\downarrow}7\%)}}$ & 78$_{\textcolor{red}{(\ensuremath{\uparrow}56\%)}}$ & 2$_{\textcolor{darkgreen}{(\ensuremath{\downarrow}67\%)}}$ & 44$_{\textcolor{darkgreen}{(\ensuremath{\downarrow}33\%)}}$ & 4$_{\textcolor{darkgreen}{(\ensuremath{\downarrow}56\%)}}$ & 73$_{\textcolor{darkgreen}{(\ensuremath{\downarrow}27\%)}}$ & 65$_{\textcolor{darkgreen}{(\ensuremath{\downarrow}20\%)}}$ & 573$_{\textcolor{darkgreen}{(\ensuremath{\downarrow}16\%)}}$\\
        & CoVe & 
        147$_{\textcolor{darkgreen}{(\ensuremath{\downarrow}36\%)}}$ & 144$_{\textcolor{red}{(\ensuremath{\uparrow}1\%)}}$ & 53$_{\textcolor{red}{(\ensuremath{\uparrow}6\%)}}$ & 2$_{\textcolor{darkgreen}{(\ensuremath{\downarrow}67\%)}}$ & 53$_{\textcolor{darkgreen}{(\ensuremath{\downarrow}20\%)}}$ & 9$_{\textcolor{gray}{(0\%)}}$ & 82$_{\textcolor{darkgreen}{(\ensuremath{\downarrow}18\%)}}$ & 84$_{\textcolor{red}{(\ensuremath{\uparrow}4\%)}}$ & 574$_{\textcolor{darkgreen}{(\ensuremath{\downarrow}16\%)}}$\\
        & RA & 
        185$_{\textcolor{darkgreen}{(\ensuremath{\downarrow}19\%)}}$& 115$_{\textcolor{darkgreen}{(\ensuremath{\downarrow}20\%)}}$ &	48$_{\textcolor{darkgreen}{(\ensuremath{\downarrow}4\%)}}$ & 0$_{\textcolor{darkgreen}{(\ensuremath{\downarrow}100\%)}}$ & 51$_{\textcolor{darkgreen}{(\ensuremath{\downarrow}23\%)}}$ & 8$_{\textcolor{darkgreen}{(\ensuremath{\downarrow}11\%)}}$ & 81$_{\textcolor{darkgreen}{(\ensuremath{\downarrow}19\%)}}$ & 87$_{\textcolor{red}{(\ensuremath{\uparrow}7\%)}}$ & 575$_{\textcolor{darkgreen}{(\ensuremath{\downarrow}16\%)}}$\\
        \cline{2-11}
        & \textbf{HaluCodeTran} &
        142$_{\textcolor{darkgreen}{(\ensuremath{\downarrow}38\%)}}$ & 103$_{\textcolor{darkgreen}{(\ensuremath{\downarrow}28\%)}}$ &	42$_{\textcolor{darkgreen}{(\ensuremath{\downarrow}16\%)}}$ & 0$_{\textcolor{darkgreen}{(\ensuremath{\downarrow}100\%)}}$ & 50$_{\textcolor{darkgreen}{(\ensuremath{\downarrow}24\%)}}$ & 3$_{\textcolor{darkgreen}{(\ensuremath{\downarrow}67\%)}}$ & 70$_{\textcolor{darkgreen}{(\ensuremath{\downarrow}30\%)}}$ & 77$_{\textcolor{darkgreen}{(\ensuremath{\downarrow}5\%)}}$ & 487$_{\textcolor{darkgreen}{(\ensuremath{\downarrow}29\%)}}$\\
        \midrule 
        \multirow{5}{*}{Mistral} & Vanilla &
        223 & 160 & 83 & 34 & 175 &	3 &	158 & 155 & 991\\
        & CoT & 
        218$_{\textcolor{darkgreen}{(\ensuremath{\downarrow}2\%)}}$ & 152$_{\textcolor{darkgreen}{(\ensuremath{\downarrow}5\%)}}$ & 88$_{\textcolor{red}{(\ensuremath{\uparrow}6\%)}}$ & 17$_{\textcolor{darkgreen}{(\ensuremath{\downarrow}50\%)}}$ & 94$_{\textcolor{darkgreen}{(\ensuremath{\downarrow}46\%)}}$ & 8$_{\textcolor{red}{(\ensuremath{\uparrow}167\%)}}$ & 127$_{\textcolor{darkgreen}{(\ensuremath{\downarrow}20\%)}}$ & 119$_{\textcolor{darkgreen}{(\ensuremath{\downarrow}23\%)}}$ & 823$_{\textcolor{darkgreen}{(\ensuremath{\downarrow}17\%)}}$\\
        & CoVe &
        185$_{\textcolor{darkgreen}{(\ensuremath{\downarrow}17\%)}}$ & 157$_{\textcolor{darkgreen}{(\ensuremath{\downarrow}2\%)}}$ & 80$_{\textcolor{darkgreen}{(\ensuremath{\downarrow}4\%)}}$ & 19$_{\textcolor{darkgreen}{(\ensuremath{\downarrow}44\%)}}$ & 153$_{\textcolor{darkgreen}{(\ensuremath{\downarrow}13\%)}}$ &	7$_{\textcolor{red}{(\ensuremath{\uparrow}133\%)}}$ & 134$_{\textcolor{darkgreen}{(\ensuremath{\downarrow}15\%)}}$ & 148$_{\textcolor{darkgreen}{(\ensuremath{\downarrow}5\%)}}$ & 883$_{\textcolor{darkgreen}{(\ensuremath{\downarrow}11\%)}}$\\
        & RA & 
        110$_{\textcolor{darkgreen}{(\ensuremath{\downarrow}51\%)}}$ & 82$_{\textcolor{darkgreen}{(\ensuremath{\downarrow}49\%)}}$ & 172$_{\textcolor{red}{(\ensuremath{\uparrow}107\%)}}$ & 10$_{\textcolor{darkgreen}{(\ensuremath{\downarrow}71\%)}}$ & 134$_{\textcolor{darkgreen}{(\ensuremath{\downarrow}23\%)}}$ & 6$_{\textcolor{red}{(\ensuremath{\uparrow}100\%)}}$ &	126$_{\textcolor{darkgreen}{(\ensuremath{\downarrow}20\%)}}$ & 140$_{\textcolor{darkgreen}{(\ensuremath{\downarrow}10\%)}}$ & 780$_{\textcolor{darkgreen}{(\ensuremath{\downarrow}21\%)}}$\\
        \cline{2-11}
        & \textbf{HaluCodeTran} &
        135$_{\textcolor{darkgreen}{(\ensuremath{\downarrow}39\%)}}$ & 93$_{\textcolor{darkgreen}{(\ensuremath{\downarrow}42\%)}}$ & 76$_{\textcolor{darkgreen}{(\ensuremath{\downarrow}8\%)}}$ & 14$_{\textcolor{darkgreen}{(\ensuremath{\downarrow}59\%)}}$ & 109$_{\textcolor{darkgreen}{(\ensuremath{\downarrow}38\%)}}$ &	4$_{\textcolor{red}{(\ensuremath{\uparrow}33\%)}}$ & 114$_{\textcolor{darkgreen}{(\ensuremath{\downarrow}28\%)}}$ & 125$_{\textcolor{darkgreen}{(\ensuremath{\downarrow}19\%)}}$ & 670$_{\textcolor{darkgreen}{(\ensuremath{\downarrow}32\%)}}$\\
        \bottomrule
    \end{tabular}
    }
\end{table*}

\begin{figure}
    \centering
    \includegraphics[width=\linewidth]{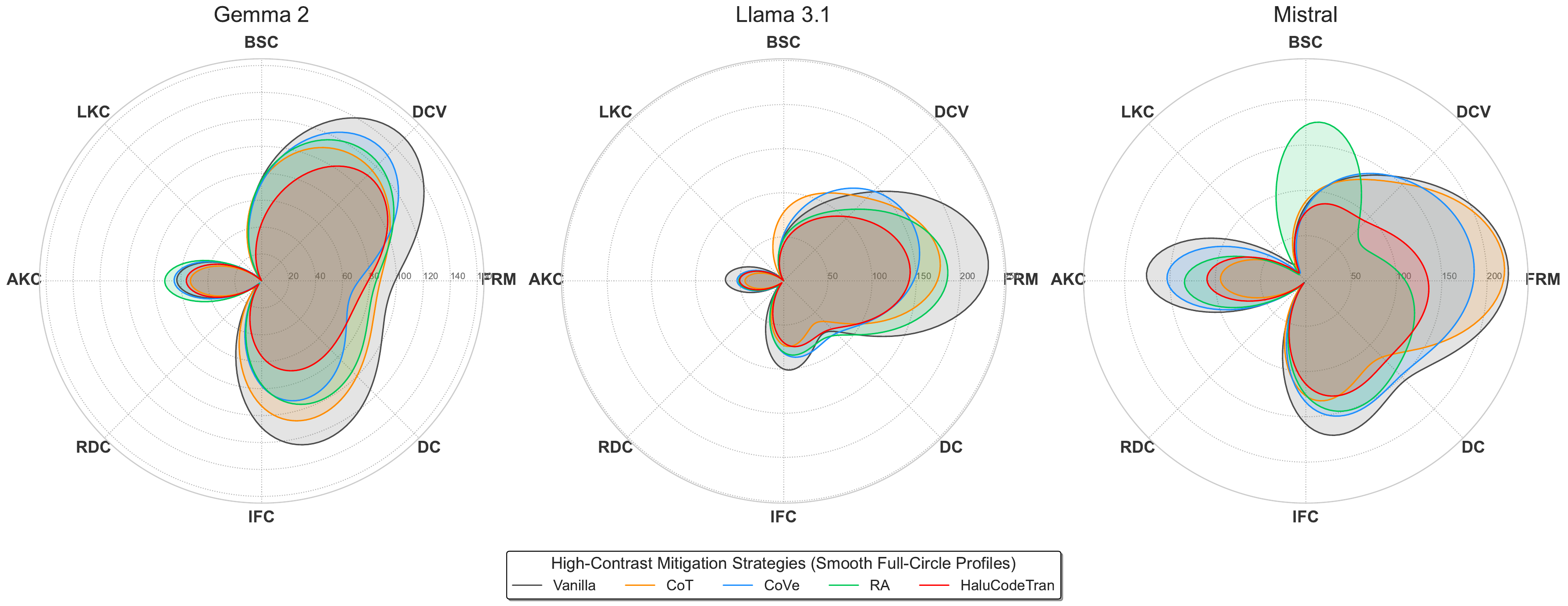}
    \caption{The hallucination distribution of different LLMs under various mitigation strategies.}
    \label{fig:distribution}
\end{figure}

\subsection{Extent of Hallucination Mitigation on Different Hallucination Types}

This subsection is the key to conducting a fine-grained and in-depth evaluation of hallucination mitigation strategies, which complements the macro performance evaluation of subsection 6.3. It quantifies how different mitigation methods reduce various hallucination subtypes across models, revealing their respective strengths and weaknesses in targeting specific hallucination types. The results provide actionable insights for the further optimization of HaluCodeTran and other mitigation frameworks, and help researchers design more refined strategies for different hallucination subtypes.

To provide a comprehensive understanding of hallucination behaviors, we conduct a fine-grained analysis of how different mitigation strategies affect various types of hallucinations across multiple models on the CodeNet dataset (Table~\ref{tab:mitigation-transport} and Figure~\ref{fig:distribution}). 
Across all models, HaluCodeTran demonstrates the most robust mitigation effect, achieving the largest average reduction in total hallucinations: 29\% for Llama 3.1, 32\% for Gemma 2, and 32\% for Mistral. 
Regarding Factual Knowledge Conflicts, especially in subtypes such as API Knowledge Conflicts (AKC) and Library Knowledge Conflicts (LKC), HaluCodeTran leads across all models; e.g., HaluCodeTran achieves a consistent reduction in AKC hallucinations across all three models (11\% reduction on Gemma 2, 24\% on Llama 3.1, 38\% on Mistral), indicating its strength in correcting incorrect or hallucinated API calls and exception handling.
Regarding Project Context Deviations, HaluCodeTran exhibits superior performance in mitigating Identifier \& Function Confusion (IFC) hallucinations, with the most significant reduction rates among all methods (49\% on Gemma 2, 30\% on Llama 3.1, 28\% on Mistral), demonstrating its ability to effectively address context deviations and inconsistent logical flow in translated code.
 
However, during hallucination mitigation by prompt engineering, the biased understanding of LLMs may lead to the emergence of new hallucinations.
For instance, RA increases BSC hallucinations in Mistral, while HaluCodeTran partially alleviates this. This suggests that certain hallucination types, such as syntax-level inaccuracies, may require targeted low-level guidance beyond the general prompting and retrieval-based techniques.

Overall, while each mitigation strategy offers strengths in specific areas, HaluCodeTran consistently delivers balanced improvements across all hallucination categories. This underscores the importance of integrating hallucination-aware constraints during both training and inference to address diverse failure modes in code generation.

\findingbox{
HaluCodeTran achieves the strongest average hallucination mitigation among the evaluated methods, especially for semantic conflicts and project context deviations. This result indicates that taxonomy-aware feedback is a useful direction for improving robust code translation.
}

\section{Threat Analysis}
The main internal validity threat comes from manual annotation. Code-translation hallucinations can be multi-faceted, and annotators may disagree about whether an erroneous fragment should be labeled as a semantic conflict, factual conflict, project-context deviation, or multiple categories. To mitigate this threat, two annotators with extensive Java, C++, and Python experience independently inspected the generated program, source program, execution result, compiler/runtime diagnostic, and reference solution before adjudication. 
A remaining limitation is that manual labeling still depends on human judgment, especially for semantically equivalent rewrites that differ substantially from the reference implementation.

Another internal validity threat is potential information leakage in retrieval-based mitigation. Because CodeNet, Avatar, and Transcoder contain many structurally similar programming tasks, a retrieval method may obtain examples that are too close to the evaluation query. We reduce this risk by excluding the query program from the retrieval candidate set and by evaluating all mitigation methods on the same per-model test split. Nevertheless, near-duplicate problems or same-family solutions may still inflate retrieval-based methods. We therefore treat the mitigation results as evidence of practical repair capability under the benchmark setting, rather than as proof that the framework fully generalizes to unseen project domains.

Construct validity is affected by the definition of hallucination. This work labels generated fragments as hallucinated when they introduce unsupported semantics, invalid language/library knowledge, or inconsistent project context. This definition is broader than factual hallucination in natural-language generation and overlaps with some categories of code translation error. We adopt the broader definition because code translation failures are often observable only through compilation, execution, and source-target semantic comparison. However, the taxonomy may not cover all edge cases, such as performance-only regressions, security-specific vulnerabilities, or errors that require whole-project context beyond the benchmark snippet.

External validity concerns the choice of languages, datasets, and models. Our study focuses on Java, C++, and Python over CodeNet, Avatar, and Transcoder, which are widely used but still do not represent all migration scenarios. Results may differ for less common languages, domain-specific frameworks, multi-file projects, or tasks requiring build-system migration. Model validity is also time-sensitive: we evaluate seven representative LLMs, including GPT-4o and open-source models, but newer models and specialized code agents may exhibit different hallucination patterns.
Reproducibility is limited by model nondeterminism and closed-model availability. Even with fixed prompts and decoding configurations, proprietary APIs may change over time, and stochastic decoding can alter generated code. 

\section{Conclusion}
This paper presents an empirical study of hallucinations in LLM-powered code translation. Through thematic analysis and manual annotation of 13,919 hallucination cases across seven LLMs, three benchmark datasets, and six language pairs, we establish a taxonomy that separates semantic consistency conflicts, factual knowledge conflicts, and project context deviations.
Based on the taxonomy, we propose HaluCodeTran, a mitigation framework that integrates retrieval-augmented examples with iterative reflective refinement. Experiments on CodeNet show that HaluCodeTran improves computational accuracy and reduces hallucination cases compared with representative prompting and retrieval baselines. These findings suggest that code-translation hallucination should be studied not only as a correctness failure, but also as a structured source of semantic, factual, and contextual error signals.
In future work, we plan to extend the taxonomy to multi-file migration scenarios, strengthen automated hallucination detection, and study training or alignment strategies that reduce hallucinations before inference-time repair is needed.

\bibliographystyle{ACM-Reference-Format}
\bibliography{software}

\end{document}